\documentclass[twocolumn]{aastex62}

\def\teff{\mbox{$T_{\rm eff}$}}
\def\logg{\mbox{log~{\it g}}}
\def\vmicro{\mbox{$\xi_{\rm turb}$}}
\def\kmsec{\mbox{km~s$^{\rm -1}$}}
\def\ocen{\mbox{$\omega$~Cen}}

\usepackage{epstopdf}
\usepackage{amsmath}
\usepackage{booktabs}

\received{ }
\revised{ }
\accepted{ }

\submitjournal{ApJ}

\shorttitle{Chemical Compositions of RR Lyrae Stars in $\omega$ Cen}
\shortauthors{Magurno et al.}

\begin{document}

\title{Chemical Compositions of Field and Globular Cluster RR~Lyrae Stars: II. $\omega$ Centauri\footnote{This paper includes data gathered with the 6.5 meter Magellan 
Telescopes located at Las Campanas Observatory, Chile.}}

\correspondingauthor{Davide Magurno}
\email{davide.magurno2@unibo.it}

\author{D. Magurno}
\affiliation{University of Roma Tor Vergata, Department of Physics, via della Ricerca Scientifica 1, 00133, Roma, Italy}
\affiliation{INAF Osservatorio Astronomico di Roma, via Frascati 33, 00040, Monte Porzio Catone RM, Italy}
\affiliation{University of Bologna, Department of Physics and Astronomy, via Irnerio 46, 40126, Bologna, Italy}

\author{C. Sneden}
\affiliation{Department of Astronomy and McDonald Observatory, The University of Texas, Austin, TX 78712, USA}

\author{G. Bono}
\affiliation{University of Roma Tor Vergata, Department of Physics, via della Ricerca Scientifica 1, 00133, Roma, Italy}
\affiliation{INAF Osservatorio Astronomico di Roma, via Frascati 33, 00040, Monte Porzio Catone RM, Italy}

\author{V. F. Braga}
\affiliation{Instituto Milenio de Astrof{\'i}sica, Santiago, Chile}
\affiliation{Departamento de F{\'i}sica, Facultad de Ciencias Exactas, Universidad Andr{\'e}s Bello, Fern{\'a}ndez Concha 700, Las Condes, Santiago, Chile}

\author{M. Mateo}
\affiliation{Department of Astronomy, University of Michigan, 1085 S. University, Ann Arbor, MI 48109, USA}

\author{S. E. Persson}
\affiliation{Observatories of the Carnegie Institution for Science, 813 Santa Barbara Street, Pasadena, CA 91101, USA}

\author{G. Preston}
\affiliation{Observatories of the Carnegie Institution for Science, 813 Santa Barbara Street, Pasadena, CA 91101, USA}

\author{F. Th\'evenin}
\affiliation{Universit\'e de La C\^ote d'Azur, OCA, Laboratoire Lagrange CNRS, BP. 4229, 06304, Nice Cedex, France}

\author{R. da Silva}
\affiliation{SSDC, via del Politecnico snc, I-00133 Roma, Italy}

\author{M. Dall'Ora}
\affiliation{INAF Osservatorio Astronomico di Capodimonte, Salita Moiariello 16, 80131 Napoli, Italy}

\author{M. Fabrizio}
\affiliation{INAF Osservatorio Astronomico di Roma, via Frascati 33, 00040, Monte Porzio Catone RM, Italy}
\affiliation{SSDC, via del Politecnico snc, I-00133 Roma, Italy}

\author{I. Ferraro}
\affiliation{INAF Osservatorio Astronomico di Roma, via Frascati 33, 00040, Monte Porzio Catone RM, Italy}

\author{G. Fiorentino}
\affiliation{INAF Osservatorio Astronomico di Bologna, Via Ranzani 1, I-40127 Bologna, Italy}

\author{G. Iannicola}
\affiliation{INAF Osservatorio Astronomico di Roma, via Frascati 33, 00040, Monte Porzio Catone RM, Italy}

\author{L. Inno}
\affiliation{Max Planck Institute f\"ur Astronomie, K\"onigstuhl 17 D-69117, Heidelberg, Germany}

\author{M. Marengo}
\affiliation{Department of Physics and Astronomy, Iowa State University, A313E Zaffarano, Ames, IA 50010, USA }

\author{S. Marinoni}
\affiliation{SSDC, via del Politecnico snc, I-00133 Roma, Italy}

\author{P. M. Marrese}
\affiliation{SSDC, via del Politecnico snc, I-00133 Roma, Italy}

\author{C. E. Mart\'inez-V\'azquez}
\affiliation{Cerro Tololo Inter-American Observatory, National Optical Astronomy Observatory, Casilla 603, La Serena, Chile}

\author{N. Matsunaga}
\affiliation{Department of Astronomy, The University of Tokyo, 7-3-1 Hongo, Bunkyo-ku, Tokyo 113-0033, Japan}

\author{M. Monelli}
\affiliation{IAC - Instituto de Astrofisica de Canarias, Calle Via Lactea, E38200 La Laguna, Tenerife, Espana}

\author{J. R. Neeley}
\affiliation{Department of Physics, Florida Atlantic University, Boca Raton, FL 33431, USA}

\author{M. Nonino}
\affiliation{INAF Osservatorio Astronomico di Trieste, Via G.B. Tiepolo 11, I-34143 Trieste, Italy}

\author{A. R. Walker}
\affiliation{Cerro Tololo Inter-American Observatory, National Optical Astronomy Observatory, Casilla 603, La Serena, Chile}

\begin{abstract}

We present a detailed spectroscopic analysis of RR Lyrae (RRL) variables 
in the globular cluster NGC~5139 (\ocen). We collected optical (4580--5330 \AA), high resolution 
($R \sim$~34,000), high signal-to-noise ratio ($\sim$200) spectra for 113 RRLs with the 
multi-fiber spectrograph M2FS at the {\it Magellan}/Clay Telescope at Las Campanas Observatory. 
We also analysed high 
resolution ($R \sim$~26,000) spectra for 122 RRLs collected with FLAMES/GIRAFFE at the 
VLT, available in the ESO archive. The current sample doubles the literature abundances of cluster and 
field RRLs in the Milky Way based on high resolution spectra. Equivalent 
width measurements were used to estimate atmospheric parameters, iron, and abundance ratios for 
$\alpha$ (Mg, Ca, Ti), iron peak (Sc, Cr, Ni, Zn), and s-process (Y) elements. We confirm 
that \ocen\ is a complex cluster, characterised by a large spread in the iron content: 
$-$2.58~$\leq$~[Fe/H]~$\leq$~$-$0.85. We estimated the average cluster abundance as 
$\langle$[Fe/H]$\rangle$~=~$-$1.80~$\pm$~0.03, with $\sigma$~=~0.33~dex. Our findings 
also suggest that two different RRL populations coexist in the cluster. The former is more 
metal-poor ([Fe/H]~$\lesssim$~$-$1.5), with almost solar abundance of Y.  
The latter is less numerous, more metal-rich, and yttrium enhanced ([Y/Fe]~$\gtrsim$~0.4).  
This peculiar bimodal enrichment only shows up in the s-process element, and it is not 
observed among lighter elements, whose [X/Fe] ratios are typical for Galactic globular clusters.

\end{abstract}


\section{Introduction} \label{sec:intro}

\ocen\ (NGC~5139) is the most massive cluster in the Galaxy \citep[4.05~$\times$~10$^6$ M$_\odot$,][]{d'souza13},
containing $\sim$1.7~$\times$~10$^6$ stars \citep{castellani07}.
\ocen\ is known to host stars that cover a broad range in metallicity,
from [Fe/H]~$\sim -2.5$ to
[Fe/H]~$\sim 0.0$ \citep{calamida09,johnson10,marino11,pancino11b,villanova14}.
This large metallicity spread, coupled with an age spread of
$\sim$2~Gyr \citep{villanova14}, suggests that \ocen\
should be identified as the remnant core of a larger pristine dwarf galaxy,
successively accreted by the Milky Way \citep{bekki03,dacosta08,marconi14,ibata19}.
On the other hand, many studies suggest different origins, 
with \ocen\ as the possible result of successive merging of inhomogeneous, 
coeval, proto-cluster clouds \citep{tsujimoto03}, 
or the result of a self-enrichment history within the cluster itself \citep{cunha02}.
A general consensus about this peculiar cluster has not been reached.

Despite the uncertainties about its origin, \ocen\ has several advantages
related to its peculiar characteristics.
Its huge number of stars permits estimation of its distance
with multiple techniques, such as variable stars like 
Miras \citep{feast65}, SX~Phoenicis \citep{mcnamara00}, 
Type II Cepheids \citep{matsunaga06}, and RR Lyraes \citep{braga18,bono19},
the tip of the red giant branch \citep{bono08b}, 
or the white dwarf cooling sequence \citep{calamida08}.
Among them, the large population of candidate RRLs \citep[$\sim$200 stars,][]{navarrete15,braga18},
makes \ocen\ the ideal laboratory for a large investigation with multi-object spectroscopy.
The multiple possibilities for a distance estimates, the large metallicity spread, 
and the large number of stars,
provide a unique possibility to calibrate RRL period-luminosity-metallicity (PLZ) 
and period-Wesenheit-metallicity (PWZ) relations with a high level of accuracy, 
which then can be applied to other RRL samples in the Galaxy. 

Photometric investigations concerning the RRLs 
in \ocen\ date back to more than one century ago \citep{bailey02} 
and they have been crucial objects for understanding the pulsation and evolutionary 
properties of old, low-mass helium burning variables 
\citep{martin51,baade58b,sandage81a,sandage81b,bono01,bono03c}.  
Optical time series CCD data were collected both by OGLE \citep{udalski92}
and by CASE \citep{kaluzny04} experiments, 
and more recently by \cite{weldrake07}. 
More recently, a complete optical \citep[$UBVRI$,][]{braga16} and near-infrared 
\citep[$JHK_s$,][]{navarrete15,braga18} census have been published. 
As usual, the high-resolution spectroscopic investigations lag
when compared to the photometric ones. Some
abundance analyses have been performed on the \ocen\ RRLs, based either on
spectroscopic \citep[][18 stars]{gratton86}, 
on spectrophotometric \citep[][131 stars]{rey00}, 
or on photometric \citep[][170 stars]{bono19}
techniques. However, the only large investigation based on high resolution spectroscopy
was performed by \citet[][74 stars collected at $R$~$\sim$~22,500]{sollima06b}.
This work aims at improving the sample of available high resolution spectroscopic
abundances for the RRLs in \ocen, based on the techniques already applied 
in \cite[][hereinafter Paper~I]{magurno18} for the smaller mono-metallic globular cluster NGC~3201.

We describe the collected dataset and the instrument settings in Section~\ref{sec:instr}. 
Section~\ref{sec:veloc} describes the analysis of radial velocities.
The investigation methodology is presented in Section~\ref{sec:abund},
and the abundance results are shown in Section~\ref{sec:iron} for iron, 
in Section~\ref{sec:alpha} for the $\alpha$-elements,
in Section~\ref{sec:heavy} for the iron-peak elements, 
and in Section~\ref{sec:heavy2} for the yttrium.
Finally, conclusions are presented in Section~\ref{sec:fine}.

\section{Instrument and data sample} \label{sec:instr}

\begin{figure}
\plotone{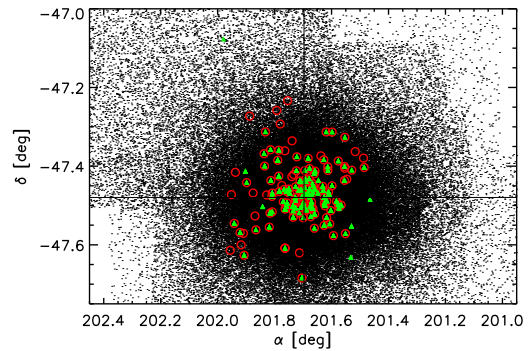}
\caption{Radial distribution of the RRLs in our spectroscopic samples. The targets
collected with M2FS and FLAMES/GIRAFFE are marked with red circles and green triangles, respectively. The cluster center is marked by the black crossing lines.  \label{fig:coord}}
\end{figure}

\begin{figure*}
\plotone{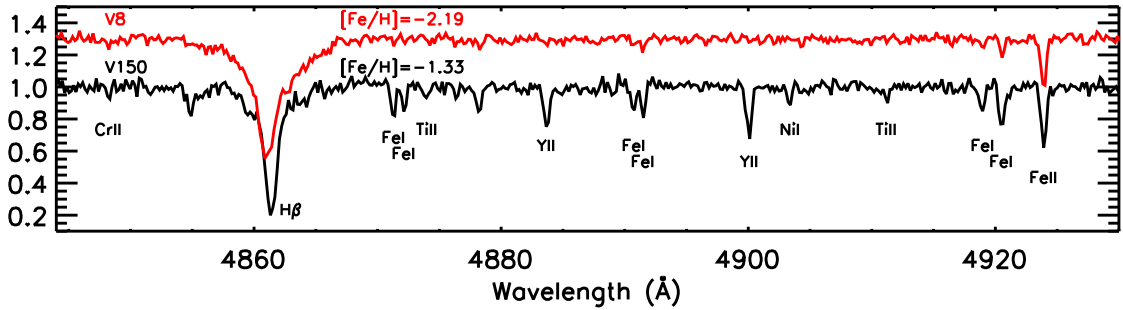}
\caption{Comparison of a portion of the M2FS spectral range for two RRab stars, V8 and V150, observed at similar phases ($\phi \sim 0.2$). The location of some useful absorption lines are marked for iron, $\alpha$, iron peak, and s-process elements. The metal-poor spectrum (red) is vertically shifted for convenience. \label{fig:spectra}}
\end{figure*}

Between 2015 February and April, we collected single epoch, 
high signal-to-noise ratio (S/N~$\sim$~200), high-resolution spectra
of 126 stars in the globular cluster \ocen\ (details in Table~\ref{tab:stars}),
uniformly distributed around the cluster center 
within a radius of about 15~arcmin from the cluster center (Figure~\ref{fig:coord}).
The spectra were collected with the Michigan/{\it Magellan} 
Fiber System (M2FS; \citealt{mateo12}) installed
at the {\it Magellan}/Clay 6.5m telescope at Las Campanas Observatory in Chile.
The selected spectrograph configuration limits the spectral coverage to 
11 overlapping echelle orders in the range 4580--5330~\AA.
The 95~$\micron$ slit size allows a
spectral resolution $R$~$\equiv$ $\lambda/\Delta\lambda$ $\simeq$~34,000.
Figure~\ref{fig:spectra} compares a portion of the M2FS 
spectral range for two RRab stars with different metallicity, collected at similar 
pulsation phases.

The sample of RRLs to be observed was selected as follows: 
we started with the variable stars catalogue by \cite{samus09}, 
and the two large RRL catalogues by \cite{kaluzny04} and \cite[][and following updates\footnote{\url{http://www.astro.utoronto.ca/~cclement/read.html}}]{clement01},
restricted to those stars within the field of view of M2FS. 
The \citeauthor{clement01} on-line database is not independent of the other two. 
We also have our own positions from the FourStar \citep{persson13} dataset, 
already used in \cite{braga18}. 
These have high accuracy because the pixel scale of FourStar is 0.16$''$/pixel 
and the typical image full width at half maximum (FWHM) 
for the infrared photometry is $\sim$0.5$''$ or better. 
We first checked the targets positions because the M2FS fibers are placed in 
pre-drilled holes that must be accurate. 
We started with a sample of all the relevant stars, weeded out the ones for which 
\citeauthor{clement01} has doubts, used the FourStar images to delete crowded stars, 
and adopted the FourStar positions where appropriate. 
This sample contained 160 RRLs, with roughly equal numbers of RRab and RRc.
The sample was divided into nine slices, each containing roughly 160/9~$\simeq$~18 stars. 
Because 16 spectra were obtained per setup (i.e. slice) 
on the two M2FS camera/detector units, 
this ensured that not all the stars could be observed. 
A choice necessitated by observing convenience and total available telescope time. 
Stars in the last slice were observed in 2016, about a year after the other eight. 
Their spectra were of inferior quality and were not included in the analysis. 
This left 8 slices $\times$ 16 stars/slice = 128 spectra. 
Of these, two spectra were unusable leaving the final sample of 126.

Unfortunately, while the sample was being cleaned,
a few radial velocity non-cluster members and light curve non-RRL
stars were mistakenly included.
Among the 126 collected spectra, three objects were marked as non cluster members
because of their almost null radial velocity, not compatible with
the cluster (see Section~\ref{sec:veloc}), 
and they were removed from the final sample.
The remaining 123 spectra can be distinguished into 113 RRLs and 10 non-RRL stars.
Only one RR Lyrae, V38, was observed twice.
The main body of the paper only refers to the RRLs, whereas the non-RRL stars are 
briefly described in the Appendix.

In addition to our M2FS data, we also analysed a sample of
560 multi-epoch spectra for 122 RRLs
from the ESO archive\footnote{Based on observations collected at the European Southern Observatory under ESO programmes 074.B-0170(A),
074.B-0170(B), 082.D-0424(A), 081.D-0255(A).}, collected with the multi-object, medium-high 
resolution spectrograph FLAMES/GIRAFFE \citep{pasquini02}.
We selected from the archive all the available RRL spectra collected with
the HR13 grism, covering the wavelength range 6120--6405~\AA\ with a spectral resolution $R \simeq$~26,400.

In total, 22 RRLs were only observed with M2FS, 
31 RRLs were only observed with GIRAFFE, 
and 91 RRLs have spectra collected with both M2FS and GIRAFFE.

\begin{deluxetable*}{lccccclccc}
\tablecaption{Photometric parameters and radial velocities for the sample stars in \ocen, collected with M2FS. \label{tab:stars}}
\tablewidth{0pt}
\tablehead{
\colhead{ID} &
\colhead{$\alpha$} &
\colhead{$\delta$} &
\colhead{period\tablenotemark{a}} &
\colhead{HJD} &
\colhead{phase} &
\colhead{type\tablenotemark{a,b}} &
\colhead{$\langle$V$\rangle$\tablenotemark{a}} &
\colhead{$A_V$\tablenotemark{a}} &
\colhead{$RV$} \\
\colhead{ } &
\colhead{J2000} &
\colhead{J2000} &
\colhead{days} &
\colhead{2450000$+$} &
\colhead{ } &
\colhead{ } &
\colhead{mag} &
\colhead{mag} &
\colhead{\kmsec}
}
\startdata
V4    &  13:26:12.94 & $-$47:24:19.2  &   0.62731846  &  7077.79451  &  1.00    &   RRab     &  14.467  &  1.119  & 202.5   \\
V5    &  13:26:18.34 & $-$47:23:12.8  &   0.51528002  &  7077.79451  &  0.59    &   RRab*    &  14.702  &  0.852  & 242.8   \\
V7    &  13:27:01.04 & $-$47:14:00.1  &   0.71303420  &  7086.82140  &  0.52    &   RRab     &  14.594  &  0.950  & 242.8   \\
V8    &  13:27:48.43 & $-$47:28:20.6  &   0.52132593  &  7125.83064  &  0.25    &   RRab     &  14.671  &  1.263  & 221.9   \\
V10   &  13:26:07.01 & $-$47:24:37.0  &   0.37475609  &  7077.79451  &  0.34    &   RRc      &  14.505  &  0.421  & 249.8   \\
V11   &  13:26:30.56 & $-$47:23:01.9  &   0.56480650  &  7087.78075  &  0.44    &   RRab*    &  14.476  &  0.453  & 243.3   \\
V12   &  13:26:27.19 & $-$47:24:06.6  &   0.38677657  &  7084.78661  &  0.58    &   RRc      &  14.498  &  0.438  & 229.3   \\
V16   &  13:27:37.71 & $-$47:37:35.0  &   0.33019610  &  7125.83063  &  0.73    &   RRc      &  14.558  &  0.487  & 228.7   \\
V18   &  13:27:45.07 & $-$47:24:56.9  &   0.62168636  &  7125.83064  &  0.87    &   RRab     &  14.551  &  1.152  & 225.8   \\
V20   &  13:27:14.05 & $-$47:28:06.8  &   0.61558779  &  7082.75671  &  0.36    &   RRab     &  14.540  &  1.098  & 232.5   \\
\enddata
\tablenotetext{a}{Reference: \cite{braga16,braga18}}
\tablenotetext{b}{The asterisks mark candidate Blazhko RRLs}
\tablenotetext{}{(This table is available in its entirety in machine-readable form.)}
\end{deluxetable*}

\section{Radial Velocities} \label{sec:veloc}

Estimating the radial velocity ($RV$) is a common way to 
establish whether a star is a globular cluster member.
\ocen\ has had many $RV$ membership investigations thanks
to its huge stellar population.
Recently, \cite{an17} estimated a cluster average velocity of
232.7~$\pm$~0.6~\kmsec, with a dispersion $\sigma$~=~14.4~\kmsec, 
by using 581 red giant branch (RGB) stars.
A decade earlier,
\cite{reijns06} performed the largest
investigation of \ocen, estimating an average radial velocity of 
231.3~$\pm$~0.3~\kmsec\ ($\sigma$~=~11.7~\kmsec), with 1589 RGB stars.
This very large cluster $RV$ makes it unlikely
that a field star
in its sightline could be erroneously identified as a cluster member.

Nevertheless, we are dealing with variable stars and 
single epoch measurements 
are affected by intrinsic radial velocity variations along the pulsation cycle.
Indeed, RRL pulsation cycles cause variations up to $\sim$70~\kmsec\
in the observed $RV$s for RRab and up to $\sim$45~\kmsec\ for RRc.
Therefore, it was necessary 
to correct their observed radial velocities for the pulsational components, in 
order to determine their systemic (cluster) velocities,  
applying the velocity templates described in the following.

We first measured the instantaneous radial velocities using 
the task {\it fxcor} in IRAF \citep{tody86,tody93}.\footnote{
IRAF is distributed by the National Optical Astronomy Observatories, which 
are operated by the Association of Universities for Research in Astronomy, 
Inc., under cooperative agreement with the National Science Foundation.}
The individual spectra were cross-correlated with a synthetic spectrum
generated with the driver {\it synth} of the local thermodynamic equilibrium (LTE) 
line analysis code MOOG\footnote{
\url{http://www.as.utexas.edu/~chris/moog.html}}
\citep{sneden73}.
This model spectrum was 
computed with the atmospheric parameters typical of 
stars in the RRL domain \citep[\teff~=~6500~K, \logg~=~2.5, 
\vmicro~=~3.0~km~s$^{-1}$, {[}Fe/H{]}~=~$-$1.5;][]{for11,sneden17} 
and then smoothed to the M2FS or GIRAFFE resolution.
The individual velocities are listed in the last column of Table~\ref{tab:stars}, 
and we assume an average error for the entire sample of $\sim$1.3~\kmsec,
as given by {\it fxcor}.

\begin{figure}
\plotone{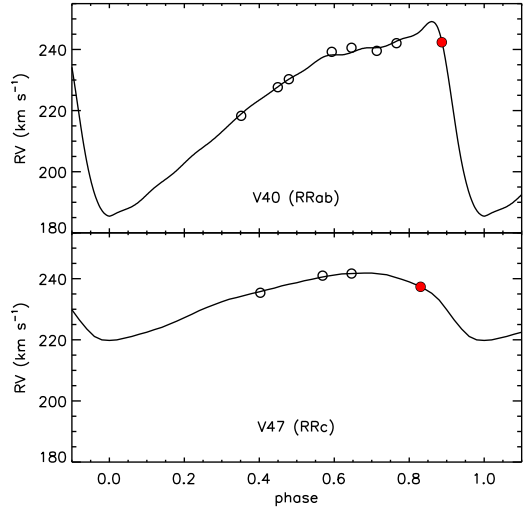}
\caption{Results of the minimization procedure to phase multiple radial velocity measurements. The template curves by \citealt{sesar12} (RRab, top panel) and by \citealt{sneden17} (RRc, bottom panel) are used as a reference to phase GIRAFFE (open black circles) and M2FS (filled red circles) observations. \label{fig:fitvel}}
\end{figure}

The use of multiple $RV$ measurements
allows us to improve the phasing of the individual data.
The phase of the individual measurements was computed by using the 
period and the epoch of maximum light, relying on the work by
\cite{braga16,braga18} for the most updated and homogeneous 
photometry, in the $UBVRIJHK_s$ bands, of the \ocen\ RRLs.
However, this approach is prone to possible systematics in cases of a large time 
interval between photometric and spectroscopic observations. 
Indeed, small errors in the determination of the period and/or in the epoch of maximum 
light could transform into large errors in the phase determination. 
Note that typical RRL periods range over about 6--18 hours (0.25--0.75 d). 
Moreover, for RRLs that are located in the cluster outskirts, we still lack accurate 
epoch of maximum light \citep{navarrete15,braga18}. On the other hand,
radial velocities are measured with high precision  
and have no dependence on photometry.
Therefore, we can use $RVs$ to compute more precise phases of the individual data points.
To do that, we first defined two radial velocity templates, for RRab and RRc stars.
\cite{sesar12} identified a linear relation between the photometric 
$V$ band amplitude ($A_{V}$, mag) and the $RV$ pulsation 
amplitude ($A_{rv}$, \kmsec) for the RRab stars.
Thus, we adopted his Equation 2

\begin{equation}
A_{\text{rv,RRab}} = 25.6(\pm 2.5) A_V + 35.0(\pm 2.3)
\end{equation}

\noindent to scale his radial velocity curve template
at the specific amplitude of each RRab star in our sample.
The same approach was applied to the RRc stars, 
from the photometry and $RVs$ presented
by \cite{sneden17}. We used the data in their Table~1 to define 
an average ratio between the velocity amplitude and the photometric 
$V$ band amplitude

\begin{equation}
A_{\text{rv,RRc}} = 54.7(\pm 2.7) A_V
\end{equation}

\noindent and we scaled their radial velocity curve template accordingly.
The next step was to fix the relative phases between the 
multiple $RV$ measurements for a single star,
according to their epochs and to the period.
Finally, we used a minimization procedure
with two free parameters (phase and average
template velocity) and two fixed ones (measured $RV$ and template amplitude)
to phase our data.
Figure~\ref{fig:fitvel} shows the alignment of the measured $RV$ points 
with the $RV$ template curves after the minimization procedure, for a RRab (top panel) 
and a RRc (bottom panel) star.
The higher the number of points, the higher the precision of the result.
We applied this method to all the stars for which at least three $RVs$ were available
(113 stars),
and we used the usual method of maximum light epoch for the remaining 
ones (31 stars),
using the most updated epochs for the RRLs in \ocen\ estimated by \cite{braga16,braga18}.

\begin{figure}
\plotone{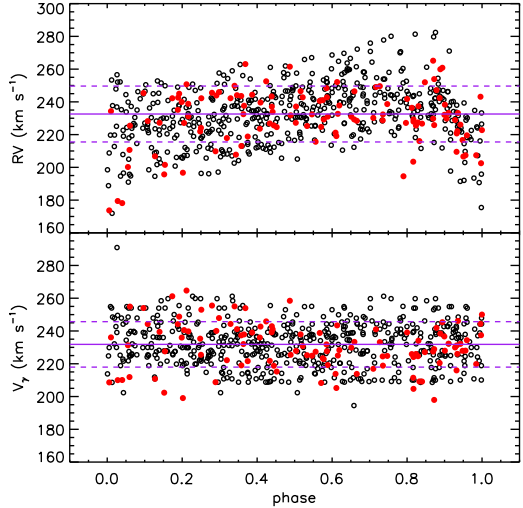}
\caption{Top panel: instantaneous radial velocity vs. phase for all the RRLs in the M2FS (red filled circles) and in the GIRAFFE (black open circles) samples. Average velocity and $1\sigma$ dispersion are shown with purple lines.
Bottom panel: as on top, but for the systemic velocity $V_\gamma$.  \label{fig:velocity}}
\end{figure}

The top panel of Figure~\ref{fig:velocity} shows the dependence on 
phase of the instantaneous radial velocity
for both the M2FS (red filled circles) and GIRAFFE (black open circles)
data sets.
Pulsational velocity effects are easily seen in this
panel.  For the individual stars, we applied the template velocity curves to 
remove these effects and derive the systemic velocities ($V_\gamma$).
For the stars with three or more $RV$ measurement, we computed $V_\gamma$ as the integral average
of the fitting template computed before.
For the other stars, we anchored the template curve, scaled to the appropriate amplitude, 
to our single measured radial velocity and phase, 
based on the epoch of maximum light, and we computed the integral average 
velocity on the template curve.
The bottom panel of Figure~\ref{fig:velocity} shows 
that the estimated $V_\gamma$ is almost independent of phase,
within the natural star-to-star scatter.
The average cluster velocity, from the joint samples of M2FS and GIRAFFE 
instantaneous velocities, 
was estimated as 232.6~$\pm$~0.7~\kmsec, with
a dispersion $\sigma$~=~17.1~\kmsec.
Once the template is applied, the average cluster velocity based on $V_\gamma$
is slightly reduced to 231.8~$\pm$~0.5~\kmsec, with a dispersion $\sigma$~=~13.9~\kmsec.
This value is in very good agreement with the cluster velocities
found by \cite{reijns06} and \cite{an17}.

\section{Abundance analysis} \label{sec:abund}

We used an equivalent width (EW) analysis method to derive
atmospheric parameters, metallicities and relative abundances 
from the M2FS sample spectra.

\subsection{Methodology} \label{sec:howdone}

We selected the 140 atomic transitions listed in Table~\ref{tab:righe}
from a collection of laboratory measurements and reverse solar analysis.
This set of lines includes all of the transitions used
in Paper~I (see their Table~3 and references therein), 
augmented by some other lines
that are detectable in the more metal-rich RRLs of \ocen.
We measured the EWs of these lines by means of a 
multi-gaussian fitting performed with the pyEW code developed by 
M. Adamow.\footnote{
\url{https://github.com/madamow/pyEW}}
Highly asymmetric lines were discarded, as well as too weak
(EW$\le$15 m\AA) or too strong 
(EW$\ge$180 m\AA) lines.
The measurement error $\epsilon$ on the EW, 
for each absorption line, can be estimated using the relation by \cite{venn12}

\begin{equation}
\epsilon = (S/N)^{-1} \times \sqrt{1.5 \times FWHM \times \delta x} + 0.1 \times EW
\end{equation}

\noindent where $\delta x$ is the pixel size of the instrument (180 m\AA).
We obtained an average error for the entire sample $\epsilon$~$\simeq$~8~m\AA.
As a final step, we used the LTE line analysis code MOOG, 
implemented in the 
Python wrapper pyMOOGi\footnote{\url {https://github.com/madamow/pymoogi}}
\citep{adamow17},
to estimate atmospheric parameters (\teff, \logg, \vmicro, 
[Fe/H]\footnote{We adopted the standard notation,
[X/H]=$A(X)-A_\sun(X)$, where $A(X)=\log(N_X)-12$. Solar abundances
refer to \cite{asplund09} within the text.}) and some relative abundances, 
using models interpolated from a grid 
of $\alpha$-enhanced ($+$0.4 in the log) atmospheres \citep{castelli03}.\footnote{
\url{http://kurucz.harvard.edu/grids.html}}
The effective temperature was estimated by minimizing 
the dependence of the abundances on the excitation potential (EP), for the individual Fe~{\sc i} lines.
The surface gravity was estimated by forcing the balance between the 
neutral and the ionized iron line abundances.
Finally, the microturbulence was estimated by minimizing the dependence of the 
abundances on the reduced equivalent width, 
RW~$\equiv$ log(EW/$\lambda$),
for the individual Fe~{\sc i} lines.

\begin{deluxetable}{cccc}
\tablecaption{Line list and atomic parameteres. \label{tab:righe}}
\tablewidth{0pt}
\tablehead{
\colhead{$\lambda$} &	
\colhead{Species} &
\colhead{EP} &
\colhead{log($gf$)}  \\
\colhead{(\AA)} &	
\colhead{ } &
\colhead{(eV)} &
\colhead{(dex)}
}
\startdata
4702.991  &    Mg {\sc i}   &   4.346  &   $-$0.44    \\     
5172.684  &    Mg {\sc i}   &   2.712  &   $-$0.39    \\     
5183.604  &    Mg {\sc i}   &   2.717  &   $-$0.17    \\     
5265.556  &    Ca {\sc i}   &   2.523  &   $-$0.26    \\     
5031.021  &    Sc {\sc ii}  &   1.357  &   $-$0.40    \\    
5239.813  &    Sc {\sc ii}  &   1.456  &   $-$0.77    \\    
4981.731  &    Ti {\sc i}   &   0.848  &    $+$0.57    \\     
4999.503  &    Ti {\sc i}   &   0.825  &    $+$0.32    \\     
5064.653  &    Ti {\sc i}   &   0.048  &   $-$0.94    \\     
5173.743  &    Ti {\sc i}   &   0.000  &   $-$1.06    \\     
\enddata
\tablenotetext{}{References:
Mg~{\sc i}, NIST database \citep{kramida18},
Ca~{\sc i}, NIST,
Sc~{\sc ii}, NIST, 
Ti~{\sc i}, \citep{lawler13}, 
Ti~{\sc ii}, \citep{wood13}, 
Cr~{\sc i}, \citep{sobeck07}, 
Cr~{\sc ii}, \citep{lawler17},
Fe~{\sc i}, \citep{obrian91,denhartog14,ruffoni14,belmonte17},
Fe~{\sc ii}, NIST,
Ni~{\sc i}, \citep{wood14},
Zn~{\sc i},  VALD database \citep{ryabchikova15},
Y~{\sc ii},  \citep{biemont11}.\\
(This table is available in its entirety in machine-readable form.)}
\end{deluxetable}

\begin{deluxetable}{lccc} 
\tablecaption{Errors on iron abundances associated with errors on the paramater estimates. \label{tab:err_param}} 
\tablewidth{0pt} 
\tablehead{ 
\colhead{Species} & 
\colhead{$\Delta$\teff} & 
\colhead{$\Delta$\logg} & 
\colhead{$\Delta$\vmicro} \\ 
\colhead{ } & 
\colhead{($\pm$500 K)} & 
\colhead{($\pm$0.5 dex)} & 
\colhead{($\pm$0.5 km s$^{-1}$)}
} 
\startdata 
$\Delta$[Fe {\sc i}/H]  & $\pm$0.35 & $\mp$0.01 & $\mp$0.04 \\ 
$\Delta$[Fe {\sc ii}/H] & $\pm$0.10 & $\pm$0.17 & $\mp$0.07 \\ 
\enddata 
\end{deluxetable} 

Errors in estimating the atmospheric parameters also reflect in the estimated
abundances. 
Table~\ref{tab:err_param} shows the effects on iron abundance 
due to typical atmospheric 
variations occurring along the entire pulsation cycle of a RRL star \citep{for11,sneden17}.
Effective temperature and surface gravity are the main sources of 
uncertainty for Fe~{\sc i} and Fe~{\sc ii}, respectively,
whereas the impact of microturbulence is relatively small.

\subsection{Metallicity scale calibration}\label{sec:scalecalib}

\begin{figure}
\plotone{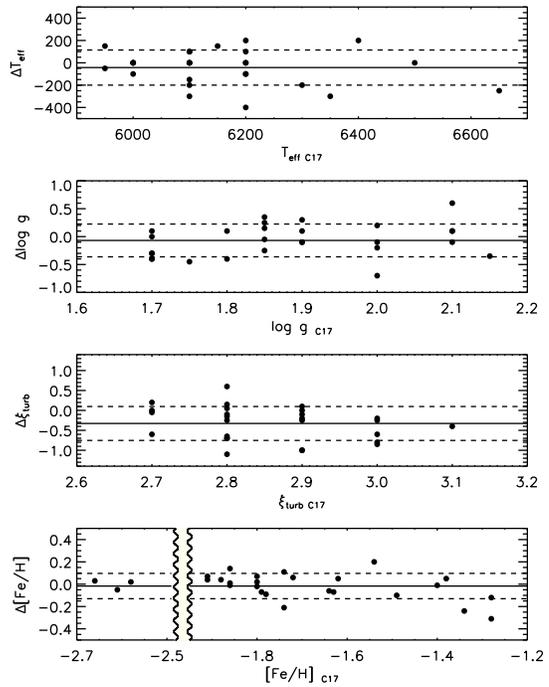}
\caption{Difference in atmospheric parameters and iron abundance between the calibrating sample by C17 (limited to the M2FS spectral range) and our estimates for the same sample, as a function of metallicity ($\Delta$X = X$_{\rm C17}$ $-$ X$_{\rm M2FS}$). The mean and 1$\sigma$ are shown with solid and dashed lines. The vertical wavy lines in the last panel break the plot for convenience, since there are no objects in the interval $-2.5 \lesssim$ [Fe/H] $\lesssim -2.0$.   \label{fig:dupont}}
\end{figure}

This study and Paper~I represent the first use of M2FS, 
with its limited spectral coverage, in a traditional abundance analysis 
of RRL stars.
It is important to understand how the metallicity scale from our analysis 
compares with previous studies.
To accomplish this, we used spectroscopic data from the high-resolution 
study of field RRLs recently reported by \citet[][hereinafter C17]{chadid17}.
They collected thousands of spectra for a sample of 35 field RRab stars,
with the du~Pont telescope at Las Campanas Observatory, over several years.
Their spectra cover a very large spectral interval, in the range 3400--9000 \AA,
much larger than the included M2FS spectral range, 
with a spectral resolution  $R$~$\simeq$~27,000.
We performed our analysis on a selection of 27 stacked spectra (S/N~$\sim$~100) by C17,
by using only the selected iron lines in the M2FS spectral range.
In Table~\ref{tab:calib}, we list the model parameters
from C17 and from our M2FS analysis, along with the 
offsets between the two metallicity estimates.
The agreement in the parameter sets is excellent: 
defining $\Delta$X~$\equiv$ X$_{\rm C17}$~$-$~X$_{\rm M2FS}$, we found
$\langle\Delta$\teff$\rangle$~=~$-$43~K ($\sigma$~=~157~K), 
$\langle\Delta$\logg$\rangle$~=~$-$0.07 ($\sigma$~=~0.29),
$\langle\Delta$\vmicro$\rangle$~=~$-$0.33~\kmsec\ ($\sigma$~=~0.43~\kmsec), and
$\langle\Delta$[Fe/H]$\rangle$~=~$-$0.02 ($\sigma$~=~0.11).
Most importantly, our M2FS-based Fe abundances are in very good 
agreement with the values obtained from the more comprehensive
spectra of C17 (see Figure~\ref{fig:dupont}).
A small offset can be noticed only for two out of the three C17 most metal-rich 
spectra, however, the differences are within 3$\sigma$ from the mean. The 
difference of the third spectrum is still within 1$\sigma$.
We can conclude that we are working on the same metallicity scale.

An additional calibration was performed in Paper~I, 
in which the same kind of analysis, 
applied to the RRLs in the monometallic globular cluster NGC~3201, 
gave comparable results with previous studies based on non-variable red giant stars.

\begin{deluxetable*}{ccccccccccc}
\tablecaption{Calibrating stars and estimated parameters. \label{tab:calib}}
\tablewidth{0pt}
\tablehead{
\colhead{ } &	
\colhead{ } &	
\multicolumn{4}{c}{C17} &
\multicolumn{4}{c}{M2FS} &
\colhead{ } \\
\cmidrule(lr){3-6} \cmidrule(lr){7-10}
\colhead{Star} &	
\colhead{phase} &
\colhead{\teff} &
\colhead{\logg} &
\colhead{\vmicro} &
\colhead{[Fe/H]} &
\colhead{\teff} &
\colhead{\logg} &
\colhead{\vmicro} &
\colhead{[Fe/H]} &
\colhead{$\Delta$[Fe/H]\tablenotemark{a}} \\
\colhead{ } &	
\colhead{ } &	
\colhead{K} &
\colhead{cgs} &
\colhead{\kmsec} &
\colhead{dex} &
\colhead{K} &
\colhead{cgs} &
\colhead{\kmsec} &
\colhead{dex} &
\colhead{dex}
}
\startdata
DN Aqr    & 0.247 & 6100 &  1.80 &  3.00 &  $-$1.78 &  6400 & 2.20 &  3.80 &  $-$1.69 & $-$0.09	\\
DN Aqr    & 0.366 & 6100 &  1.80 &  2.80 &  $-$1.74 &  6000 & 1.70 &  3.45 &  $-$1.85 & +0.11	\\
SW Aqr    & 0.280 & 6500 &  1.90 &  2.90 &  $-$1.40 &  6500 & 2.00 &  2.90 &  $-$1.39 & $-$0.01	\\
SW Aqr    & 0.413 & 6200 &  2.00 &  2.90 &  $-$1.34 &  6600 & 2.70 &  3.90 &  $-$1.10 & $-$0.24	\\
X Ari     & 0.301 & 6200 &  1.90 &  2.80 &  $-$2.66 &  6300 & 2.00 &  3.50 &  $-$2.69 & +0.03	\\
X Ari     & 0.374 & 6100 &  2.15 &  2.80 &  $-$2.61 &  6300 & 2.50 &  3.90 &  $-$2.56 & $-$0.05	\\
X Ari     & 0.470 & 6000 &  1.90 &  2.80 &  $-$2.58 &  6000 & 1.80 &  3.05 &  $-$2.60 & +0.02	\\
RR Cet    & 0.335 & 6100 &  1.70 &  2.90 &  $-$1.49 &  6250 & 2.10 &  3.15 &  $-$1.39 & $-$0.10	\\
RR Cet    & 0.554 & 5950 &  1.70 &  3.10 &  $-$1.63 &  6000 & 2.10 &  3.50 &  $-$1.56 & $-$0.07	\\
SX For    & 0.308 & 6000 &  1.70 &  2.70 &  $-$1.79 &  6100 & 2.00 &  2.75 &  $-$1.72 & $-$0.07	\\
SX For    & 0.363 & 6000 &  1.70 &  2.80 &  $-$1.80 &  6000 & 1.70 &  2.90 &  $-$1.78 & $-$0.02	\\
SX For    & 0.454 & 5950 &  1.70 &  2.80 &  $-$1.80 &  5800 & 1.60 &  2.95 &  $-$1.87 & +0.07	\\
V Ind     & 0.323 & 6400 &  2.00 &  2.70 &  $-$1.54 &  6200 & 1.80 &  2.70 &  $-$1.74 & +0.20	\\
V Ind     & 0.396 & 6200 &  2.00 &  2.80 &  $-$1.64 &  6300 & 2.20 &  2.65 &  $-$1.58 & $-$0.06	\\
V Ind      & 0.471 & 6200 &  2.10 &  2.70 &  $-$1.62 &  6100 & 2.00 &  2.50 &  $-$1.67 & +0.05	\\
SS Leo    & 0.314 & 6200 &  2.10 &  2.90 &  $-$1.86 &  6200 & 2.20 &  3.10 &  $-$1.87 & +0.01	\\
SS Leo    & 0.410 & 6100 &  2.10 &  2.80 &  $-$1.88 &  6100 & 2.00 &  3.50 &  $-$1.92 & +0.04	\\
SS Leo    & 0.557 & 6000 &  1.90 &  2.90 &  $-$1.91 &  6000 & 1.60 &  3.90 &  $-$1.98 & +0.07	\\
ST Leo    & 0.217 & 6650 &  2.00 &  3.00 &  $-$1.28 &  6900 & 2.10 &  3.20 &  $-$0.97 & $-$0.31	\\
ST Leo    & 0.316 & 6300 &  1.70 &  2.70 &  $-$1.28 &  6500 & 2.00 &  3.30 &  $-$1.16 & $-$0.12	\\
ST Leo    & 0.452 & 6150 &  2.10 &  2.80 &  $-$1.38 &  6000 & 1.50 &  2.75 &  $-$1.43 & +0.05	\\
VY Ser    & 0.229 & 6200 &  1.85 &  2.90 &  $-$1.91 &  6200 & 2.10 &  2.80 &  $-$1.95 & +0.04	\\
VY Ser    & 0.293 & 6200 &  1.85 &  2.90 &  $-$1.86 &  6000 & 1.60 &  3.00 &  $-$2.00 & +0.14	\\
VY Ser    & 0.366 & 6100 &  1.85 &  2.80 &  $-$1.86 &  6100 & 1.70 &  2.20 &  $-$1.85 & $-$0.01	\\
W Tuc     & 0.284 & 6350 &  1.75 &  3.00 &  $-$1.74 &  6650 & 2.20 &  3.85 &  $-$1.53 & $-$0.21	\\
W Tuc     & 0.397 & 6100 &  1.85 &  3.00 &  $-$1.72 &  6000 & 1.50 &  3.25 &  $-$1.78 & +0.06	\\
W Tuc     & 0.475 & 6100 &  1.85 &  3.00 &  $-$1.80 &  6100 & 1.90 &  3.60 &  $-$1.82 & +0.02	\\
\enddata
\tablenotetext{a}{$\Delta$[Fe/H] = [Fe/H]$_{\text{C17}} -$ [Fe/H]$_{\text{M2FS}}$}
\end{deluxetable*}

\subsection{Stellar Parameters}\label{sec:wcenparams}

\begin{figure}
\plotone{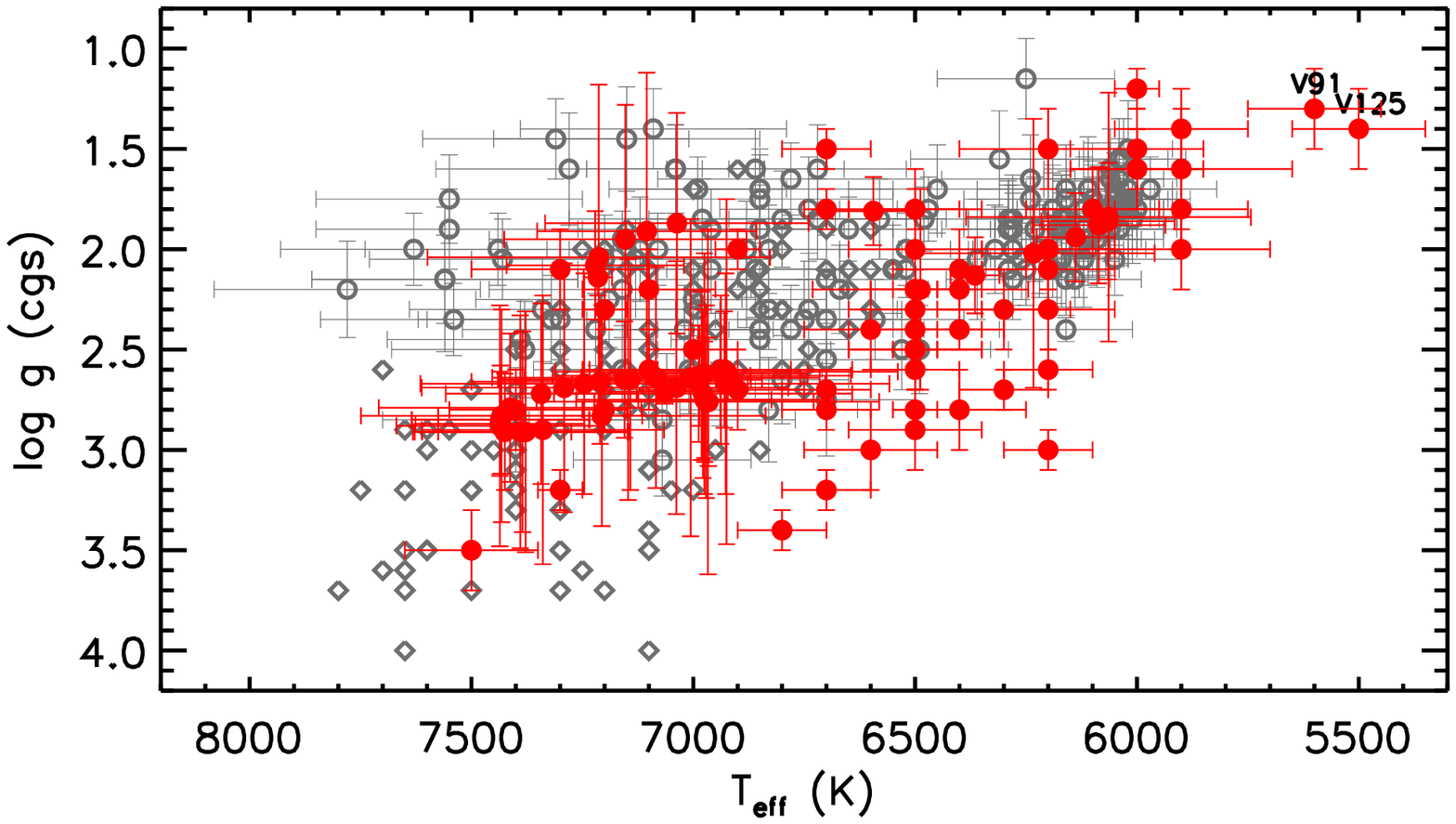}
\caption{\teff\ vs \logg\ estimated with the EW method for our cluster RRLs (filled red circles) and a sample of literature values for field RRab \citep[][open black circles]{for11} and field RRc \citep[][open black diamonds]{sneden17}. Note that the axis orientation is reversed, to resemble the structure of a HR diagram. \label{fig:teff_logg}}
\end{figure}

A total of 58 M2FS \ocen\ spectra (57 objects) showed enough useful 
lines to perform a full spectroscopic parameter determination and  abundance analysis.
Figure~\ref{fig:teff_logg} compares the relation \teff--\logg\ for 
our M2FS sample (filled red circles), 
with the parameters obtained by \cite{for11} and \cite{sneden17} for field RRLs (open black marks).     
The agreement of the two samples is good, with a few exceptions.
In particular, two stars (V91 and V125) appear cooler than the bulk of the data.
However, a visual inspection of the spectra does not give any argument to reject
these stars as non-RRLs, so they are kept in the sample.

\begin{figure}
\plotone{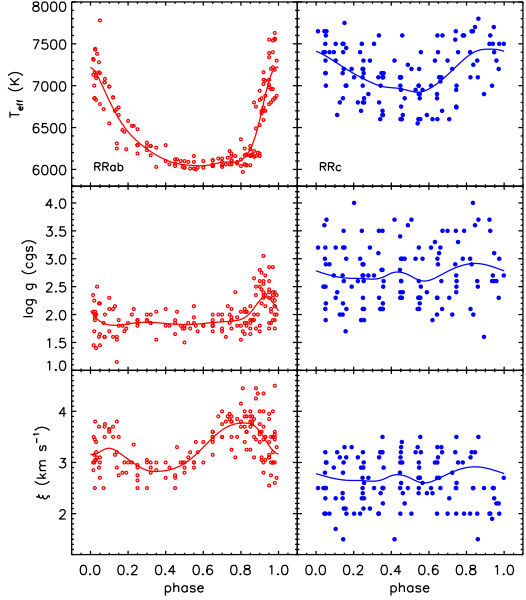}
\caption{Atmospheric parameters vs. phase for RRab \citep[][left panels]{for11} and RRc \citep[][right panels]{sneden17}. Polynomial fits on the different samples are shown with solid lines (see text for more details).  \label{fig:average}}
\end{figure}

Another 51 M2FS spectra did not have enough iron lines
to retrieve reliable atmospheric parameters with the EW method.
The S/N ratio of the spectra is quite homogeneous. The lack of lines is caused 
either by the low metallicity of the target, or to a hotter pulsation phase, or 
both.
In particular, many of them did not have enough measurable 
Fe~{\sc i} lines to estimate 
effective temperature from Boltzmann excitation equilibrium, and others did 
not have any Fe~{\sc ii} lines to estimate surface gravity from Saha ionization
equilibrium.
However, we were able to estimate average parameters starting from their phase.
\cite{for11} analysed 11 field RRab, covering their entire pulsation cycles with multiple observations,
showing that the atmospheric parameters have a relatively slow and regular
variation along the pulsation cycle.
The same applies to the 19 RRc analysed by \cite{sneden17}. 
However, the two quoted groups show, at fixed
pulsation phase, a significant difference in the spread for which
we do not have yet an explanation (see Figure~\ref{fig:average}).
Both the samples were collected at the du~Pont telescope and
were analysed with the same approach adopted by C17. 
This guarantees that we are still in the same calibration system as shown
in Section~\ref{sec:scalecalib}.
We applied the PEGASUS (PEriodic GAuSsian Uniform and Smooth fit)
procedure described by \cite{inno15b} to fit the  
atmospheric parameter distributions
as a function of phase (solid lines in Figure~\ref{fig:average}).
This was applied to the two individual samples of RRab and RRc,
to obtain phase average parameters (hereinafter called PAP) to be used in 
the abundance determinations.
The fitting function is in the form

\begin{equation}
y(\phi)=A_0+\sum_{i=1}^{N}A_i\ e^{-B_i \sin^2(\pi (\phi-\Phi_i)) }
\end{equation}

\begin{deluxetable*}{lcccccc}
\tablecaption{Polynomial coefficients of the atmospheric parameter fitting functions. \label{tab:fitpar}}
\tablewidth{0pt}
\tablehead{
\colhead{ } &
\multicolumn{3}{c}{RRab} &
\multicolumn{3}{c}{RRc} \\
\cmidrule(lr){2-4}
\cmidrule(lr){5-7}
\colhead{coeff.} &
\colhead{\teff} &
\colhead{\logg} &
\colhead{\vmicro} &
\colhead{\teff} &
\colhead{\logg} &
\colhead{\vmicro}
}
\startdata
$N$      &    5             &  3             &  6             &  5             &  5             &  4            \\
$A_0$    &    106019        &  1.89243       &  2.71275       &  14383.4       &  2.58097       &  2.55349      \\
$A_1$    &    $-$489.403    &  $+$0.431386   &  $+$0.620255   &  $-$440.029    &  $+$0.335342   &  $+$0.103329  \\
$A_2$    &    $-$98093.6    &  $-$0.0834223  &  $-$0.0470036  &  $-$7261.09    &  $-$0.243433   &  $-$0.209276  \\
$A_3$    &    $-$3760.66    &  $-$0.0719636  &  $+$0.21354    &  $-$55.011     &  $+$0.305311   &  $+$0.0910824 \\
$A_4$    &    $-$813.769    &  {\ldots}      &  $+$0.332713   &  $-$26.8128    &  $+$0.0202035  &  $+$0.19808   \\
$A_5$    &    $-$1071.11    &  {\ldots}      &  $+$0.594971   &  $-$16.9569    &  $+$0.0280391  &  {\ldots}     \\
$A_6$    &    {\ldots}      &  {\ldots}      &  $+$0.192962   &  {\ldots}      &  {\ldots}      &  {\ldots}     \\
$B_1$    &       19.3373    &     16.5656    &     3.17812    &     2.17685    &     2.2762     &     5.32572   \\
$B_2$    &       0.0309723  &     7.78696    &     18.4264    &     0.0714892  &     7.73026    &     9.36127   \\
$B_3$    &       0.943952   &     3.36144    &     17.8028    &     13.8483    &     9.66563    &     12.0717   \\
$B_4$    &       10.3451    &  {\ldots}      &     4.28953    &     15.0278    &     30.0616    &     9.83077   \\
$B_5$    &       3.40488    &  {\ldots}      &     4.6093     &     32.1722    &     65.5694    &  {\ldots}     \\
$B_6$    &    {\ldots}      &  {\ldots}      &     18.6277    &  {\ldots}      &  {\ldots}      &  {\ldots}     \\
$\Phi_1$ &       0.893882   &     0.932017   &     0.685243   &     0.649517   &     0.843868   &     0.398946  \\
$\Phi_2$ &       0.500645   &     0.121129   &     0.292199   &     0.325824   &     0.525006   &     0.947203  \\
$\Phi_3$ &       0.0963617  &     0.500979   &     0.892172   &     0.54464    &     0.477258   &     0.759689  \\
$\Phi_4$ &       0.819111   &  {\ldots}      &     0.14965    &     0.360766   &     0.234945   &     0.558797  \\
$\Phi_5$ &       0.716676   &  {\ldots}      &     0.825808   &     0.922283   &     0.402275   &  {\ldots}     \\
$\Phi_6$ &    {\ldots}      &  {\ldots}      &     0.0919317  &  {\ldots}      &  {\ldots}      &  {\ldots}     \\
\enddata
\end{deluxetable*}

\noindent where $y$ is one of the atmospheric parameters
(\teff, \logg, \vmicro) and $\phi$ is the pulsation phase. All the coefficients are 
provided in Table~\ref{tab:fitpar}.
Unfortunately, the errors based on this approach are about one order
of magnitude larger than those based on a EW analysis, for two reasons.

i) The atmospheric parameters are the average ones, and their standard deviations
can be as high as $\sigma_{T_{\text{eff}}} \simeq$~400~K, 
$\sigma_{\log g} \simeq$~0.6~dex, $\sigma_{\xi_{\text{turb}}} \simeq$~0.6~\kmsec, especially 
for the first overtone mode and during the phases of maximum light. 
This causes uncertainties in the abundances up to
0.4--0.5~dex (see Table~\ref{tab:err_param}).

ii) The spectra are not good candidates for a full EW analysis due to the paucity of good lines. 
This means that it is more difficult to decide whether a line is good or not
with respect to the others, simply because there are few lines to compare with.
Indeed, in a group of tens of lines, an outlier is
immediately identified and removed. At the contrary, with only one or two lines it
is not possible to exclude any value.

However, this approach gives better results than an estimate of the parameters
based on photometric colors, as used, for example, by \cite{sollima06b} and \cite{johnson10}.
To confirm that, we applied both the PAP and the photometric approach 
to the sample of RRLs for which we spectroscopically estimated the atmospheric
parameters. 
For the photometric approach, we used the parametrizations defined by 
\cite{johnson10}, with the 
light curves in V and K$_s$ collected by 
\cite{braga16,braga18}. The microturbulence was defined by minimizing the
abundance dependence on the reduced EW, once fixed \teff\ and \logg.
The average differences, in terms of the atmospheric parameters,
between the spectroscopic and the photometric estimates, confirm that the 
PAP appears to be more accurate.
Indeed, defining $\Delta X_{\text{PE}} \equiv X_{\text{PAP}} - X_{\text{EW}}$
and $\Delta X_{\text{phE}} \equiv X_{\text{photometric}} - X_{\text{EW}}$,
where $X$ represents one of the atmospheric parameters,
we found $\langle \Delta \teff_{\text{,PE}} \rangle \simeq$~240~K,
$\langle \Delta \teff_{\text{,phE}}\rangle \simeq$~370~K,
$\langle \Delta \logg_{\text{PE}} \rangle \simeq$~0.04,
$\langle \Delta \logg_{\text{phE}}\rangle \simeq$~0.6,
$\langle \Delta \vmicro_{\text{,PE}} \rangle \simeq$ 0.5~\kmsec,
$\langle \Delta \vmicro_{\text{,phE}}\rangle \simeq$~0.08~\kmsec.
The parameter dispersions in the two approaches are similar,
of the order of $\sigma_{\text{\teff}} \simeq$~500~K, $\sigma_{\text{\logg}} \simeq$~0.6,
$\sigma_{\text{\vmicro}} \simeq$~0.5~\kmsec.
We then applied the PAP approach to retrieve the abundances for the additional 51 RRLs.

For the remaining four spectra, no abundance analysis was possible
because of the absence of useful lines or of phase information.

\section{Metallicity Distribution} \label{sec:iron}

\begin{figure}
\plotone{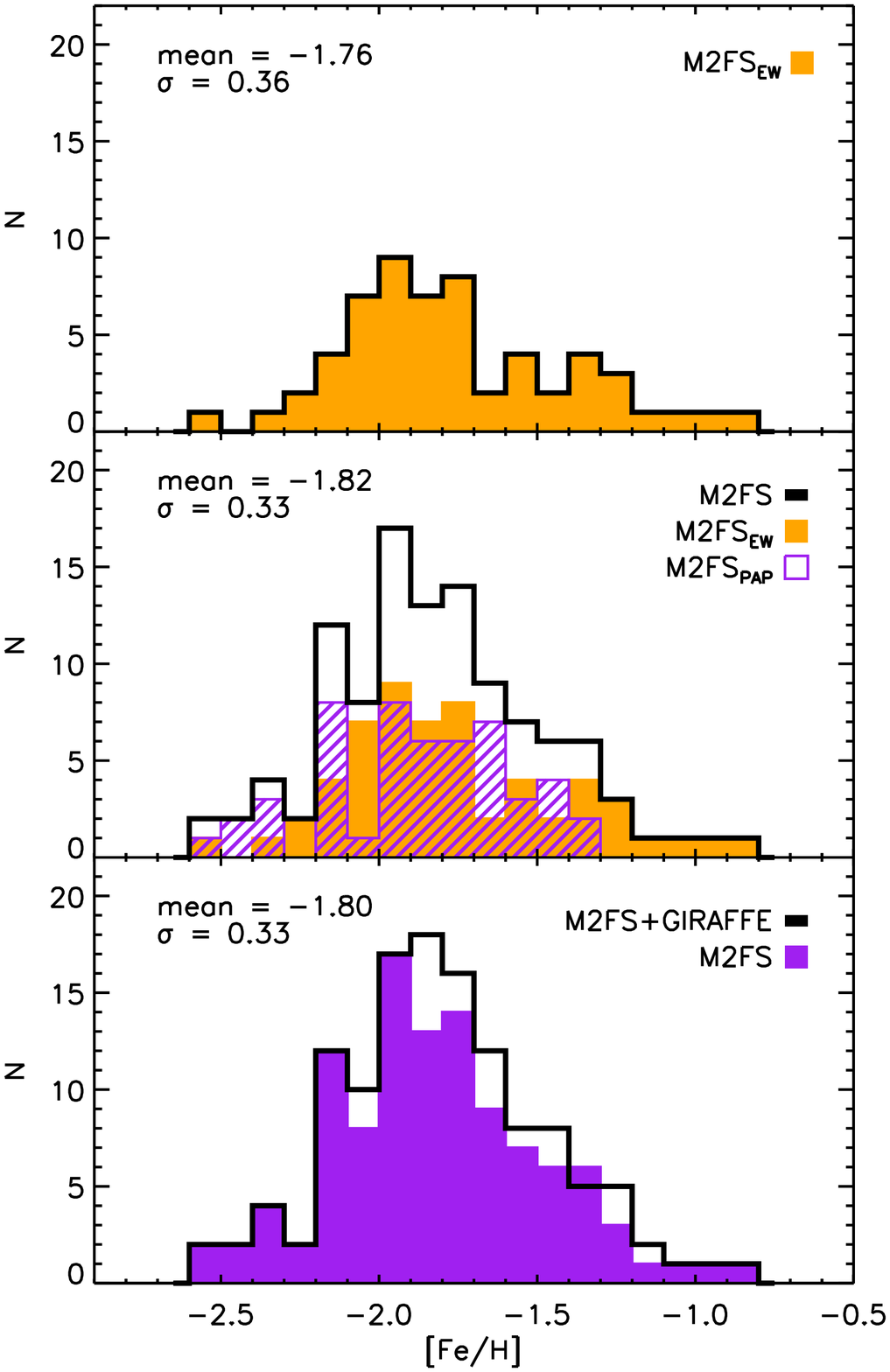}
\caption{Top panel: metallicity distribution for the M2FS sample of RRLs in \ocen, whose parameters were estimated with the EW approach.
Middle panel: metallicity distribution for the full sample of M2FS RRLs (black thick line). The orange filled area is the same as in the top panel, showing the sample estimated with the EW approach. The purple shaded area shows the sample estimated with the PAP approach. 
Bottom panel: metallicity distribution of the entire sample of RRLs collected with M2FS and GIRAFFE (black thick line). The purple filled area is the same total M2FS sample shown in the middle panel.  \label{fig:iron_hist}}
\end{figure}

The iron abundance estimates for the individual stars are 
listed in Table~\ref{tab:iron}.
The sample of 57 RRLs for which we applied a full spectroscopic analysis 
based on the EW method shows an average cluster metallicity 
$\langle$[Fe/H]$\rangle$~=~$-1.76 \pm 0.05$ and
a large star-to-star dispersion $\sigma = 0.36$, as expected for \ocen\ 
\cite{freeman75,pancino00,calamida09,bono19}.
As shown in Figure~\ref{fig:iron_hist}, top panel, our sample
has its metallicity peak at about [Fe/H]~=~$-$1.9 and a pronounced tail 
toward higher metallicities, up to [Fe/H]~=~$-$0.85.
The low metallicity tail is much less evident,
with the most metal-poor RRL estimated at [Fe/H]~=~$-$2.53.

\begin{figure}
\plotone{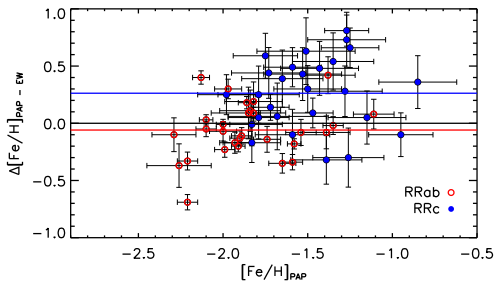}
\caption{Calibration of the metallicity scale obtained with the EW and the PAP approach for the M2FS sample RRLs. The mean difference of the two samples was used to correct the PAP sample.  \label{fig:ew-pap}}
\end{figure}

Before taking into account the 51 additional RRLs obtained with the PAP approach,
we performed a further calibration by computing, for the RRLs in the EW sample,
the corresponding iron abundances with the PAP method.
In Figure~\ref{fig:ew-pap}, we plotted the difference in [Fe/H] between the two approaches, 
for the same stars,
as a function of the iron abundance estimated with the PAP approach.
There is clearly a large spread in the points, because the average atmospheric
parameters can have higher or lower values than the real ones.
Moreover, the parametrization for the RRab appears very promising, with a
difference between the two approaches very close to zero, whereas the RRc appear, 
on average, more metal-rich with the PAP approximation.
We therefore applied a zero point calibration to the PAP sample of \ocen\ RRLs, 
according to the pulsation type,
to make it consistent with the more accurate spectroscopic one.
We also applied the same kind of correction to all the other elements, 
after performing a similar calibration based on their [X/H] abundances.
Figure~\ref{fig:iron_hist}, middle panel, shows the histogram for the entire M2FS sample 
(black thick line) after the calibration,
together with the two subsamples: the EW (orange filled area) and the PAP (purple shaded area). 
For the joint EW and PAP samples, we derived
$\langle$[Fe/H]$\rangle$~=~$-1.82 \pm 0.03$ ($\sigma$~=~0.33).
This mean value is only 0.06 dex lower than that derived
with the pure EW analysis.
Once again, the distribution peaks at about [Fe/H]~=~$-$1.9, 
with a longer metal-rich tail and a shorter metal-poor one.

Table~\ref{tab:Feab_Fec} shows the mean iron content of the two RRL populations,
RRab and RRc, with the different approaches adopted.
It can be noticed that the EW method produces very similar results for both RRab and RRc,
with a difference in the average iron content limited to 0.03~dex.
At the contrary, the PAP method produces a RRab population that is 0.13~dex more metal-rich
than the RRc one.
In particular, the RRab sample has the same average abundance ($\langle$[Fe/H]$\rangle$~=~$-$1.78) 
for both methods, whereas the RRc sample is more metal-poor in the PAP
sample ($\langle$[Fe/H]$\rangle$~=~$-$1.91) than in the EW one ($\langle$[Fe/H]$\rangle$~=~$-$1.75).
However, a lower metallicity for the PAP sample is expected, since the method
is applied to those spectra with a limited number of lines.

\begin{deluxetable*}{lccccc}
\tablecaption{Iron abundances for the considered samples of \ocen\ RRLs. Mean value, standard deviation, and number of stars for each sample are listed at the bottom. \label{tab:iron}}
\tablewidth{0pt}
\tablehead{
\colhead{ID} &
\colhead{[Fe/H]$_{\rm EW}$} &
\colhead{[Fe/H]$_{\rm PAP}$} &
\colhead{[Fe/H]$_{\rm GIRAFFE}$} &
\colhead{$n$\tablenotemark{a}} &
\colhead{[Fe/H]$_{\rm tot}$\tablenotemark{b}}
}
\startdata
V4        &  {\ldots}          &  $-$1.82 $\pm$ 0.10  &  {\ldots}         &  {\ldots}  & $-$1.82 $\pm$ 0.10  \\ 
V5        &  $-$1.40 $\pm$ 0.02  &  {\ldots}          &  $-$1.51 $\pm$ 0.29 &  2         & $-$1.40 $\pm$ 0.03  \\ 
V7        &  $-$1.76 $\pm$ 0.03  &  {\ldots}          &  {\ldots}         &  {\ldots}  & $-$1.76 $\pm$ 0.03  \\ 
V8        &  $-$2.19 $\pm$ 0.07  &  {\ldots}          &  {\ldots}         &  {\ldots}  & $-$2.19 $\pm$ 0.07  \\ 
V10       &  $-$2.23 $\pm$ 0.04  &  {\ldots}          &  {\ldots}         &  {\ldots}  & $-$2.23 $\pm$ 0.04  \\ 
V11       &  $-$1.88 $\pm$ 0.05  &  {\ldots}          &  {\ldots}         &  {\ldots}  & $-$1.88 $\pm$ 0.05  \\ 
V12       &  {\ldots}          &  $-$2.37 $\pm$ 0.04  &  {\ldots}         &  {\ldots}  & $-$2.37 $\pm$ 0.04  \\ 
V15       &  {\ldots}          &  {\ldots}          &  $-$1.68 $\pm$ 0.35 &  3         & $-$1.68 $\pm$ 0.35  \\ 
V16       &  {\ldots}          &  $-$2.00 $\pm$ 0.11  &  {\ldots}         &  {\ldots}  & $-$2.00 $\pm$ 0.11  \\ 
V18       &  {\ldots}          &  $-$1.89 $\pm$ 0.52  &  {\ldots}         &  {\ldots}  & $-$1.89 $\pm$ 0.52  \\ 
\midrule
\ocen\    & $-$1.76 $\pm$ 0.05   &  $-$1.87 $\pm$ 0.04  & $-$1.71 $\pm$  0.04 &            & $-$1.80 $\pm$ 0.03  \\
$\sigma$  &   {0.36}           &     {0.30}         &     {0.28}        &            &   {0.33}          \\
$N$       &         {57}       &       {51}         &     {44}          &            &   {125}           \\
\enddata
\tablenotetext{a}{Multiplicity of the GIRAFFE spectra.}
\tablenotetext{b}{Weighted mean on the inverse square of the measurement errors.}
\tablenotetext{}{(This table is available in its entirety in machine-readable form.)}
\end{deluxetable*}

\begin{deluxetable*}{lccccccccc}
\tablecaption{Mean iron abundance and standard deviation for the analysed samples, with distintion between different RRL pulsation modes. The number of stars for each subsample is also indicated. \label{tab:Feab_Fec}}
\tablewidth{0pt}
\tablehead{
\colhead{ } &
\multicolumn{3}{c}{M2FS$_{\text {EW}}$} &
\multicolumn{3}{c}{M2FS$_{\text {PAP}}$} &
\multicolumn{3}{c}{GIRAFFE} \\
\cmidrule(lr){2-4} 
\cmidrule(lr){5-7} 
\cmidrule(lr){8-10}
\colhead{Mode} &
\colhead{$N$} &
\colhead{$\langle$[Fe/H]$\rangle$} &
\colhead{$\sigma$} &
\colhead{$N$} &
\colhead{$\langle$[Fe/H]$\rangle$} &
\colhead{$\sigma$} &
\colhead{$N$} &
\colhead{$\langle$[Fe/H]$\rangle$} &
\colhead{$\sigma$}
}
\startdata
RRab	& 28 & $-$1.78 & 0.34   & 14 & $-$1.78 & 0.26  & 37 &   $-$1.68 & 0.28    \\
RRc	& 29 & $-$1.75 & 0.39   & 37 &  $-$1.91 & 0.31  & 7 &  $-$1.83 & 0.26    \\
\enddata
\end{deluxetable*}

\subsection{The GIRAFFE sample}

Among the $\sim$500 GIRAFFE spectra for which we measured a radial velocity,
only a limited sample of 99 spectra (44 objects, 27 in common with the M2FS sample) 
showed high enough S/N (40~$\lesssim$~S/N~$\lesssim$~110)
and useful iron lines to perform accurate EW measurements.
However, the number of iron lines was too limited for a
spectroscopic determination of the atmospheric parameters. 
In particular, they lacked useful Fe~{\sc ii} lines to balance the surface gravity.
Therefore, we applied the PAP approach to estimate the atmospheric parameters,
then the abundances, for all the stars
with available phase information and good enough iron lines.

As a first step,
we averaged the abundances for the stars with multiple GIRAFFE measurements.
Then, we compared the stars in common between the GIRAFFE and the M2FS samples.
This defined a zero point calibration for RRab and RRc, 
used to move the entire GIRAFFE sample to 
the M2FS, spectroscopic, metallicity scale. 
After the scaling, we performed a weighted averaged of the abundances for the stars 
with multiple measurements of the two spectrographs (last column of Table~\ref{tab:iron}),
assuming the inverse square of the error as weight.
We ended with a sample of 125 RRLs, 
whose distribution is shown in the bottom panel of Figure~\ref{fig:iron_hist}.
The shape of the distribution still remains essentially the same, with
$\langle$[Fe/H]$\rangle$~=~$-1.80 \pm 0.03$
and a dispersion $\sigma$~=~0.33~dex.

\begin{figure}
\plotone{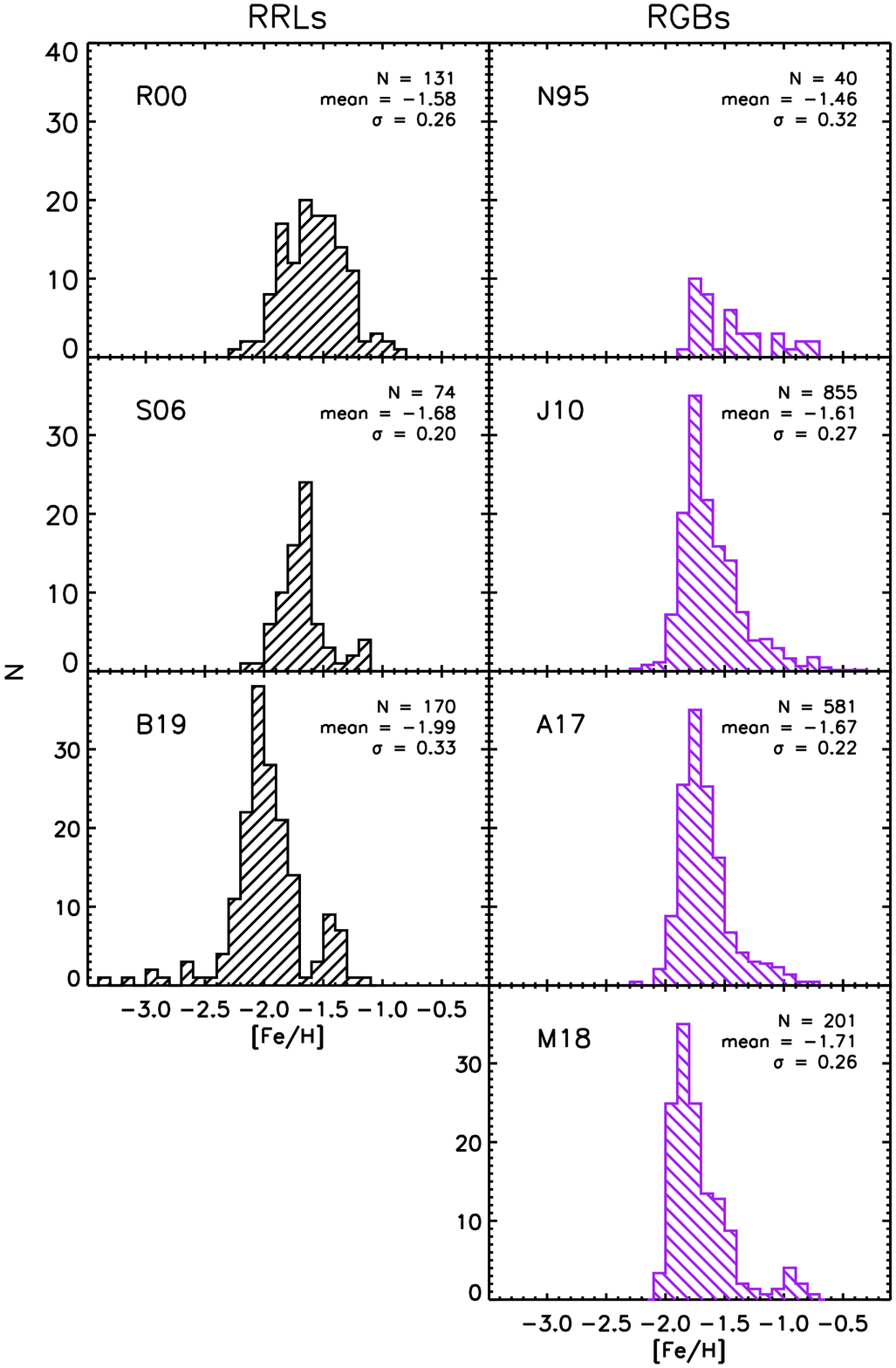}
\caption{Metallicity distribution for the sample of RRLs (black, left column) and RGBs (purple, right column) in \ocen, available in the literature (R00: \citealt{rey00}; S06: \citealt{sollima06b}; B19: \citealt{bono19}; N95: \citealt{norris95}; J10: \citealt{johnson10}; A17: \citealt{an17}; M18: \citealt{mucciarelli18,mucciarelli19}). J10, A17, and M18 have been scaled to a maximum height of 35 for plotting reasons. They should be multiplied by $\sim$5, $\sim$4, and $\sim$1.5 respectively, to obtain the real scale. The number of stars in the sample, the mean, and the standard deviation are labelled in the top right corner.  \label{fig:histograms}}
\end{figure}

\begin{figure}
\plotone{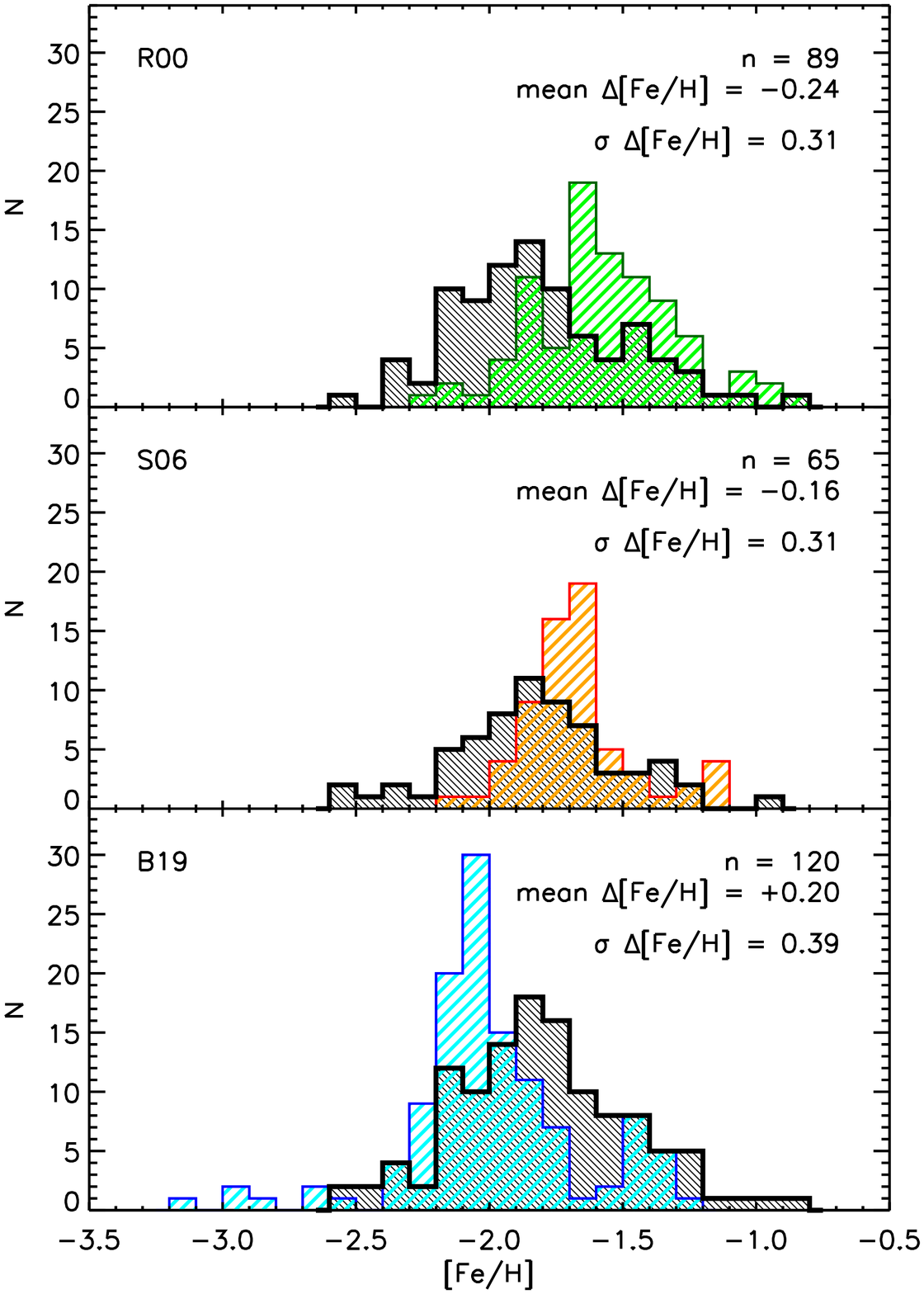}
\caption{Comparison of the metallicity distribution for our entire \ocen\ sample (M2FS+GIRAFFE, black) and the RRL samples available in the literature (R00: \citealt{rey00}; S06: \citealt{sollima06b}; B19: \citealt{bono19}). For each panel, only the stars in common ($\pmb{n}$) between the two works are considered. Mean and standard deviation of the differences among the two samples ($\Delta$[Fe/H] = this work $-$ literature) are labelled on the top right corners of each panel.  \label{fig:hist_compare}}
\end{figure}

\subsection{Comparison with the literature}

The large dispersion in the metallicity of \ocen\ is a well known 
attribute that has been investigated for decades. 
Previous studies of both RGBs \citep{norris95,johnson10,an17,mucciarelli18,mucciarelli19} and 
RRLs \citep{butler78,gratton86,rey00,sollima06b,bono19} 
clearly showed metallicity spreads between 0.20 and 0.45~dex. 
In Figure~\ref{fig:histograms}, we collected the [Fe/H] distributions for the largest 
and most recent studies.
With the exception of \citet[][hereinafter R00]{rey00}, whose distribution 
is essentially symmetric, the histograms show the longer metal-rich 
tail distribution also found in the present study.
However, the RRL based analysis of \citet[][hereinafter S06]{sollima06b} and \citet[][hereinafter B19]{bono19},
as well as the RGB analysis by \cite{mucciarelli18,mucciarelli19},
show the metal-rich tail as a separated secondary peak,
whereas the current sample shows either a metal-rich secondary peak for [Fe/H]~$\ge$~$-$1.5 
(M2FS$_{EW}$) or a well defined metal-rich shoulder (M2FS+GIRAFFE).
This was already noticed, among the others, by \cite{norris96,norris97} with Ca abundance and kinematics data,
but we will discuss this point in more detail in Section~\ref{sec:heavy2}.

It is worth mentioning that the iron distribution peak in the literature is, on average, $\sim$0.2~dex more metal-rich than our
estimate, with the exception of B19 who found a slightly more metal-poor peak.
These differences are mainly due to the techniques used to estimate the atmospheric parameters.
Indeed, S06, \cite{johnson10}, and \cite{an17} used photometrically estimated parameters
that, as already mentioned in Section~\ref{sec:wcenparams}, give slightly higher \teff\ and \logg.
Comparing our GIRAFFE analysis with the sample by S06, that is included in our sample,
we found an average difference, for the stars in common, 
$\Delta$[Fe/H]$_{\text{GIRAFFE - S06}}$~=~$-$0.13.
Since this difference can not be due to the spectra, it can only be related to the 
applied technique.
Finally, R00 used the photometric {\it hk} index to indirectly estimate the metallicity,
while B19 used a technique based on PLZ theoretical predictions.
The comparison of the literature results
with our entire \ocen\ sample (M2FS+GIRAFFE) is shown in Figure~\ref{fig:hist_compare}, 
only considering the stars in common among each work and this one.
The average differences in [Fe/H] are
$\langle\Delta$[Fe/H]$\rangle_{\text{this work--R00}}$~=~$-$0.24 (n~=~89, $\sigma$~=~0.31), 
$\langle\Delta$[Fe/H]$\rangle_{\text{this work--S06}}$~=~$-$0.16 (n~=~65, $\sigma$~=~0.31),
and $\langle\Delta$[Fe/H]$\rangle_{\text{this work--B19}}$~=~$+$0.20 (n~=~120, $\sigma$~=~0.39).

\section{The $\alpha$-elements: M\lowercase{g}, C\lowercase{a}, \lowercase{and} T\lowercase{i}} \label{sec:alpha}

\begin{figure}
\plotone{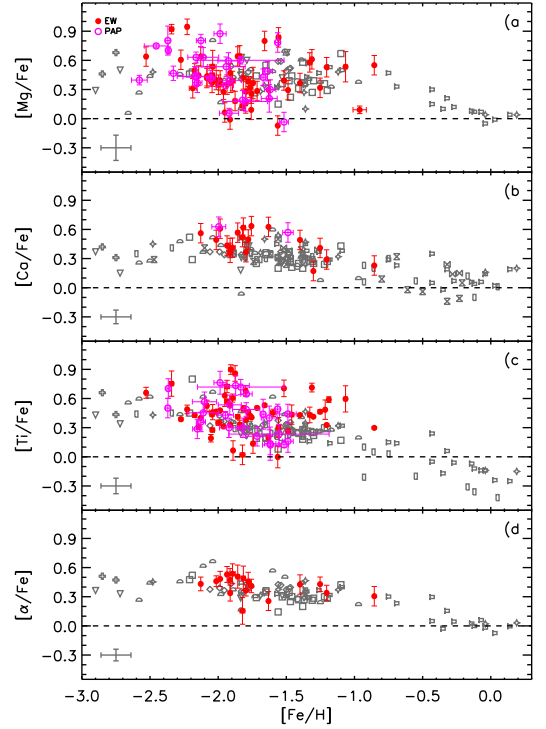}
\caption{$\alpha$-elements vs. iron abundances for the RRLs in \ocen. The stars analysed with the EW approach are marked with red filled circles, those analysed with the PAP approach are marked with magenta open circles. The black symbols mark the field halo RRLs collected with high-resolution spectroscopy from the literature. In each panel, the black error bar on bottom left corner shows the mean individual errors for the literature sample. \label{fig:wcenalfa}}
\end{figure}

\begin{figure}
\plotone{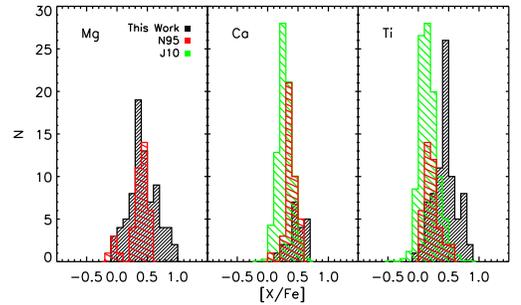}
\caption{$\alpha$-elements distribution for RGB stars in \ocen\ (N95: \citealt{norris95}; J10: \citealt{johnson10}), compared with our RRLs sample. J10 has been scaled to a maximum value of 28 for plotting reason.  \label{fig:other1}}
\end{figure}

\begin{figure}
\plotone{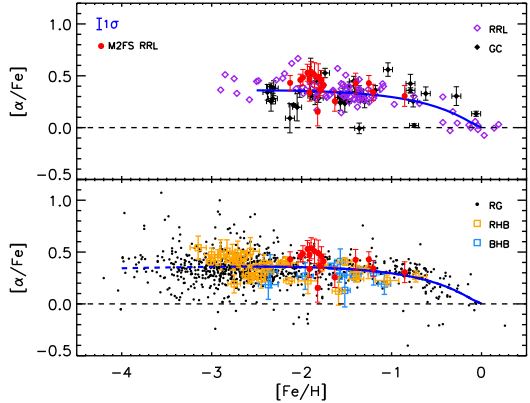}
\caption{Top panel:  $\alpha$-elements (Mg+Ca+Ti) vs. iron abundances for the individual RRLs in \ocen\ (red filled circles). The sample is compared with Galactic globulars (filled black diamonds, \citealt{pritzl05b,carretta09b,carretta10}) and field halo RRLs (open purple diamonds, same samples as in Figure~\ref{fig:wcenalfa}). The solid blue line shows the log-normal fit of the two joint samples, with the 1$\sigma$ dispersion shown by the blue bar in the top left corner. 
Bottom panel: as on top, but compared with field halo giants \citep[black dots,][]{frebel10c} and RHB--BHB field stars \citep[orange--blue squares,][]{for10}. The dashed blue line shows an extrapolation of the fit toward lower iron abundances.  \label{fig:alfafit}}
\end{figure}

\begin{figure}
\plotone{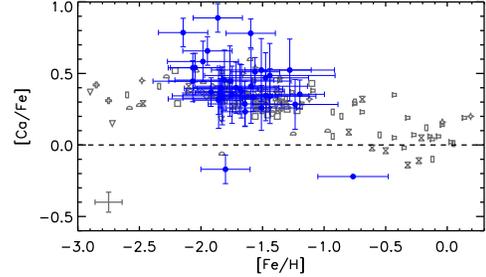}
\caption{As in Figure \ref{fig:wcenalfa}, panel (b), but for abundances derived from the GIRAFFE spectra (blue filled circles). \label{fig:cagiraffe}}
\end{figure}

\begin{deluxetable*}{lccccccccc}
\tablecaption{$\alpha$-, iron-peak, and s-process element abundances for the considered samples of \ocen\ RRLs collected with M2FS. Mean value, standard deviation, and number of stars for each species are listed at the bottom. \label{tab:elementi}}
\tablewidth{0pt}
\tablehead{
\colhead{ID} &
\colhead{[Mg/Fe]} &
\colhead{[Ca/Fe]} &
\colhead{[Ti/Fe]} &
\colhead{[ $\alpha$/Fe]\tablenotemark{a}} &
\colhead{[Sc/Fe]} &
\colhead{[Cr/Fe]} &
\colhead{[Ni/Fe]} &
\colhead{[Zn/Fe]} &
\colhead{[Y/Fe]}
}
\startdata
V4	&     0.19 $\pm$ 0.10 &      {\ldots}       &      {\ldots}         &   {\ldots}        &      {\ldots}       &      {\ldots}       &      {\ldots}       &      {\ldots}        &         {\ldots}        \\
V5	&     0.37 $\pm$ 0.10 &     0.49 $\pm$ 0.10 &     0.43 $\pm$ 0.05   &  0.43 $\pm$ 0.10  &     0.08 $\pm$ 0.10 &     0.01 $\pm$ 0.04 &     0.01 $\pm$ 0.10 &     0.30 $\pm$ 0.16  &        0.38 $\pm$ 0.08  \\
V7	&     0.26 $\pm$ 0.10 &     0.63 $\pm$ 0.10 &     0.40 $\pm$ 0.03   &  0.41 $\pm$ 0.07  &      {\ldots}       &  $-$0.28 $\pm$ 0.03 &      {\ldots}       &      {\ldots}        &        0.19 $\pm$ 0.10  \\
V8	&     0.31 $\pm$ 0.10 &      {\ldots}       &      {\ldots}         &   {\ldots}        &      {\ldots}       &     0.36 $\pm$ 0.10 &      {\ldots}       &      {\ldots}        &         {\ldots}        \\
V10	&     0.95 $\pm$ 0.14 &      {\ldots}       &     0.49 $\pm$ 0.09   &   {\ldots}        &      {\ldots}       &     0.24 $\pm$ 0.10 &      {\ldots}       &      {\ldots}        &         {\ldots}        \\
V11	&     0.18 $\pm$ 0.10 &      {\ldots}       &     0.86 $\pm$ 0.11   &   {\ldots}        &     0.66 $\pm$ 0.10 &      {\ldots}       &      {\ldots}       &      {\ldots}        &         {\ldots}        \\
V12	&     0.80 $\pm$ 0.21 &      {\ldots}       &     0.50 $\pm$ 0.10   &   {\ldots}        &     0.33 $\pm$ 0.10 &      {\ldots}       &      {\ldots}       &      {\ldots}        &         {\ldots}        \\
V16	&      {\ldots}       &     0.63 $\pm$ 0.10 &      {\ldots}         &   {\ldots}        &      {\ldots}       &      {\ldots}       &      {\ldots}       &      {\ldots}        &         {\ldots}        \\
V18	&      {\ldots}       &      {\ldots}       &      {\ldots}         &   {\ldots}        &      {\ldots}       &      {\ldots}       &      {\ldots}       &      {\ldots}        &         {\ldots}        \\  
V20	&     0.09 $\pm$ 0.10 &      {\ldots}       &     0.41 $\pm$ 0.03   &   {\ldots}        &     0.10 $\pm$ 0.10 &     0.08 $\pm$ 0.10 &     0.09 $\pm$ 0.12 &      {\ldots}        &     $-$0.10 $\pm$ 0.10  \\
\midrule
\ocen & 0.43 $\pm$ 0.03 &  0.47 $\pm$ 0.03 &  0.44 $\pm$ 0.02 &  0.41 $\pm$ 0.02 & 0.11 $\pm$ 0.04 &  0.09 $\pm$ 0.02 &  0.06 $\pm$ 0.04 &  0.30 $\pm$ 0.05 & 0.25 $\pm$ 0.05 \\
$\sigma$ & 0.22 & 0.13 & 0.19 & 0.10 & 0.21 & 0.18 & 0.17 & 0.11 & 0.31 \\
$N$ & 78 & 21 & 80 & 18 & 32 & 52 & 20 & 6 & 40\\
\enddata
\tablenotetext{a}{Biweight mean of Mg, Ca, and Ti abundances.}
\tablenotetext{}{(This table is available in its entirety in machine-readable form.)}
\end{deluxetable*}

The M2FS spectra cover a relatively short wavelength range, limiting
the $\alpha$-element line measurements to only three species: Mg, Ca, and Ti.
Titanium is not a \textquotedblleft pure\textquotedblright\ 
$\alpha$-element (its dominant isotope is $^{48}$Ti instead of 
$^{44}$Ti), however, its abundance at
low metallicity usually mimics those of the other $\alpha$-elements.
Moreover, titanium lines are the most numerous after iron in the M2FS wavelength range,
and in some cases they are the only observable ones among the $\alpha$.
Indeed, up to 13 Ti~{\sc i} and Ti~{\sc ii} lines were measured in a single spectrum, whereas
Mg~{\sc i} lines were limited to three at most, and only a single Ca~{\sc i} line was measured, if any.
We estimated their average cluster abundances for the EW sample as 
$\langle$[Mg/Fe]$\rangle$~=~0.41~$\pm$~0.03,
$\langle$[Ca/Fe]$\rangle$~=~0.46~$\pm$~0.03, and 
$\langle$[Ti/Fe]$\rangle$~=~0.44~$\pm$~0.03.
The dispersion of Mg is the largest one ($\sigma$~=~0.22),
whereas Ca has the smallest dispersion ($\sigma$~=~0.13), 
but also the lowest number of measurements,
and Ti lies in between ($\sigma$~=~0.19).
Adding the RRLs analysed with the PAP approach does not change significantly the final 
results, but they double the number of stars:
$\langle$[Mg/Fe]$\rangle$~=~0.43~$\pm$~0.03,
$\langle$[Ca/Fe]$\rangle$~=~0.47~$\pm$~0.03, and $\langle$[Ti/Fe]$\rangle$~=~0.44~$\pm$~0.02,
with exactly the same dispersions as before.

Table~\ref{tab:elementi} lists the individual star abundances, 
and Figure~\ref{fig:wcenalfa}
compares them with those of
the field halo RRLs,
collected with high-resolution (R~$\ge$~25,000) spectroscopy,
available in the literature 
\citep[][more details about the literature samples can be found in the Appendix of Paper~I]{clementini95,fernley96,lambert96,kolenberg10,for11,hansen11,liu13,govea14,pancino15,chadid17,sneden17}.
The agreement of the two considered samples is evident.
The running mean of the two groups is the same, within the dispersion,
in the metallicity range covered by the \ocen\ stars, where the $\alpha$-element
abundances are almost constant or slightly decreasing toward higher metallicities.
We also computed the [$\alpha$/Fe] abundance for the individual RRLs
as the biweight mean of the three considered element abundances 
(bottom panel of Figure~\ref{fig:wcenalfa}). 
Details about this robust iterative estimator of location can be found in \cite{beers90}.
The [$\alpha$/Fe] abundance was only estimated for those RRLs showing lines of all the
three elements. This limits the sample to 18 RRLs, 
but the homogeneity of the results is preserved.
The cluster average abundance was estimated as 
$\langle$[$\alpha$/Fe]$\rangle$~=~0.41~$\pm$~0.02 ($\sigma$~=~0.10).

Figure~\ref{fig:other1} compares our $\alpha$-element 
abundances with those derived in earlier investigations of \ocen\ red giants,
and there is a general agreement.
It can be noticed that our estimates of Mg abundances are more scattered than
those by \cite{norris95}, but with similar mean values. However, this difference
can simply be caused by the different sample size (78 vs. 40 measurements). 
On the contrary, Ca and Ti
have similar dispersions between RRLs and RGBs, but with higher abundances
for our sample, especially for Ti, with respect to both \cite{norris95} and
\cite{johnson10}.
The two different populations display quite similar 
$\alpha$-element abundances over the entire metallicity range.
To further investigate the chemical enrichment of the $\alpha$-elements in different
stellar components, we also compared our results for \ocen\ 
with similar abundances available in the literature for
Galactic globular clusters 
\citep{pritzl05b,carretta09b,carretta10}, field halo red/blue 
horizontal branch stars \cite[RHB, BHB,][]{for10}, and kinematically selected field halo red giants \citep{frebel10c}.
Figure~\ref{fig:alfafit} shows all the previous samples and a log-normal analytical fit 
of their [$\alpha$/Fe] vs. [Fe/H], computed with
the Equation~1 in Paper~I.
It is remarkable that all the observed components, which are RRLs or non-variables,
and which are cluster or field stars, agree with the fit within 1$\sigma$.
This suggests that all these components experienced very similar chemical 
enrichment histories for these $\alpha$-elements, also supporting 
a common old (t $\ge$ 10 Gyr) age for all of them.

From the GIRAFFE spectra, the only element other than Fe that was 
possible to measure with sufficient precision is Ca.
We estimated  [Ca/Fe] for 40 out of 44 stars in the sample, 
shown in Figure~\ref{fig:cagiraffe}.
Despite a few outliers, the average Ca abundance for the GIRAFFE spectra is in good 
agreement with the literature values
for field halo RRLs. We estimated the average abundance for the GIRAFFE sample, 
excluding the evident outliers with a sigma clipping procedure, 
as $\langle$[Ca/Fe]$\rangle$~=~0.42~$\pm$~0.03, with a dispersion $\sigma$~=~0.17.

\section{The Iron-peak Elements: S\lowercase{c}, C\lowercase{r}, N\lowercase{i}, \lowercase{and} Z\lowercase{n}} \label{sec:heavy}

We estimated abundances for a few heavier elements 
in the iron-peak group (Z~=~21--30): Sc, Cr, Ni, and Zn,
only observable in the M2FS spectra.
There are significant differences in the number of useful lines among 
the four species; indeed, some of these
elements are undetectable in many of the stars (Table~\ref{tab:elementi}).
Cr is the most broadly represented element, observed in about 50\%
of the RRLs ($\sim$70\% for the EW sample alone), 
with up to ten lines in the best case and at least a couple
of lines for the majority of the spectra (either Cr~{\sc i} or Cr~{\sc ii}).
On the other hand,
Zn~{\sc i} was only observed in a handful of stars, especially in the 
metal-rich tail of the sample, with only one or two lines.
Sc~{\sc ii} and Ni~{\sc i} both have few observed lines in the M2FS spectral range, but
the number of RRLs showing them is between those with Cr and Zn.
The average cluster abundances were estimated as
$\langle$[Sc/Fe]$\rangle$~=~0.11~$\pm$~0.04, 
$\langle$[Cr/Fe]$\rangle$~=~0.09~$\pm$~0.02,
$\langle$[Ni/Fe]$\rangle$~=~0.06~$\pm$~0.04, and 
$\langle$[Zn/Fe]$\rangle$~=~0.30~$\pm$~0.05.

\begin{figure}
\plotone{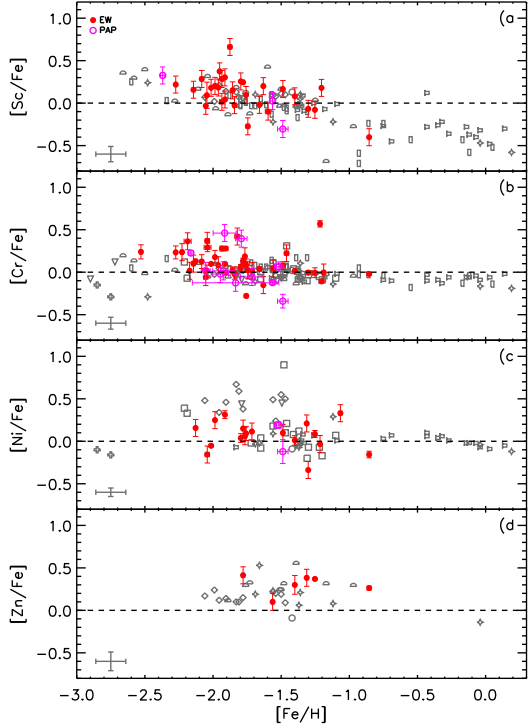}
\caption{Iron-peak elements vs. iron abundances for the RRLs in \ocen. The stars analysed with the EW approach are marked with red filled circles, those analysed with the PAP approach are marked with magenta open circles. The black symbols mark the field halo RRLs collected with high-resolution spectroscopy from the literature. In each panel, the black error bar on bottom left corner shows the mean individual errors for the literature sample.  \label{fig:heavy}}
\end{figure}

Figure~\ref{fig:heavy} shows the comparison between the 
abundances of the iron-peak elements
for the individual \ocen\ RRLs and for the field halo RRLs available in the literature
(see Appendix in Paper~I for more details about the literature sample).
The agreement of the two groups is very good. 
The running means of the two samples with metallicity
are nearly the same.
The dispersions, in the metallicity range covered by \ocen, are also very similar 
between the two groups and to those of the $\alpha$-elements,
for Sc ($\sigma_{\text{\ocen}}$~=~0.21, $\sigma_{\text{halo}}$~=~0.15), 
Cr ($\sigma_{\text{\ocen}}$~=~0.18, $\sigma_{\text{halo}}$~=~0.09), 
Ni ($\sigma_{\text{\ocen}}$~=~0.17, $\sigma_{\text{halo}}$~=~0.26), 
and Zn ($\sigma_{\text{\ocen}}$~=~0.11, $\sigma_{\text{halo}}$~=~0.14).
Once again, the chemical enrichment history of the RRLs in the Galactic halo 
appears to be similar for both field and cluster stars.

\begin{figure}
\plotone{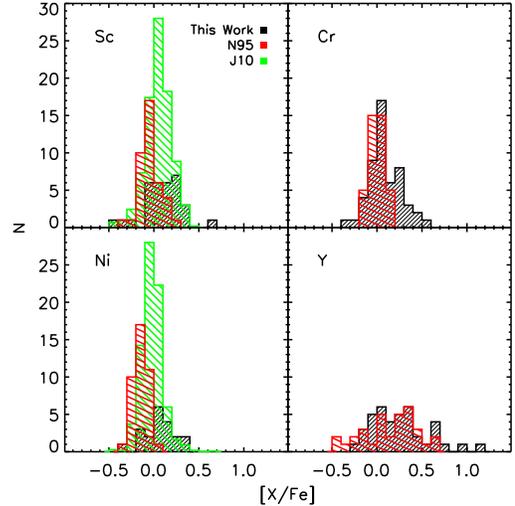}
\caption{Iron-peak and s-process elements distribution for RGB stars in \ocen\ (N95: \citealt{norris95}; J10: \citealt{johnson10}), compared with our RRLs sample. Zn is not shown because missing in the RGB samples. J10 has been scaled to a maximum value of 28 for plotting reason. \label{fig:other2}}
\end{figure}

Our Fe-group abundances from RRL stars are also in good 
agreement with those derived from the RGB samples.
The results obtained by \cite{norris95} and \cite{johnson10}
for the \ocen\ RGBs (Figure~\ref{fig:other2}) are in 
general agreement with our
RRL sample, with only limited differences in the average values, 
suggesting similar enrichment histories for the two stellar groups.

\section{The s-process Element: Y} \label{sec:heavy2}

\begin{figure}
\plotone{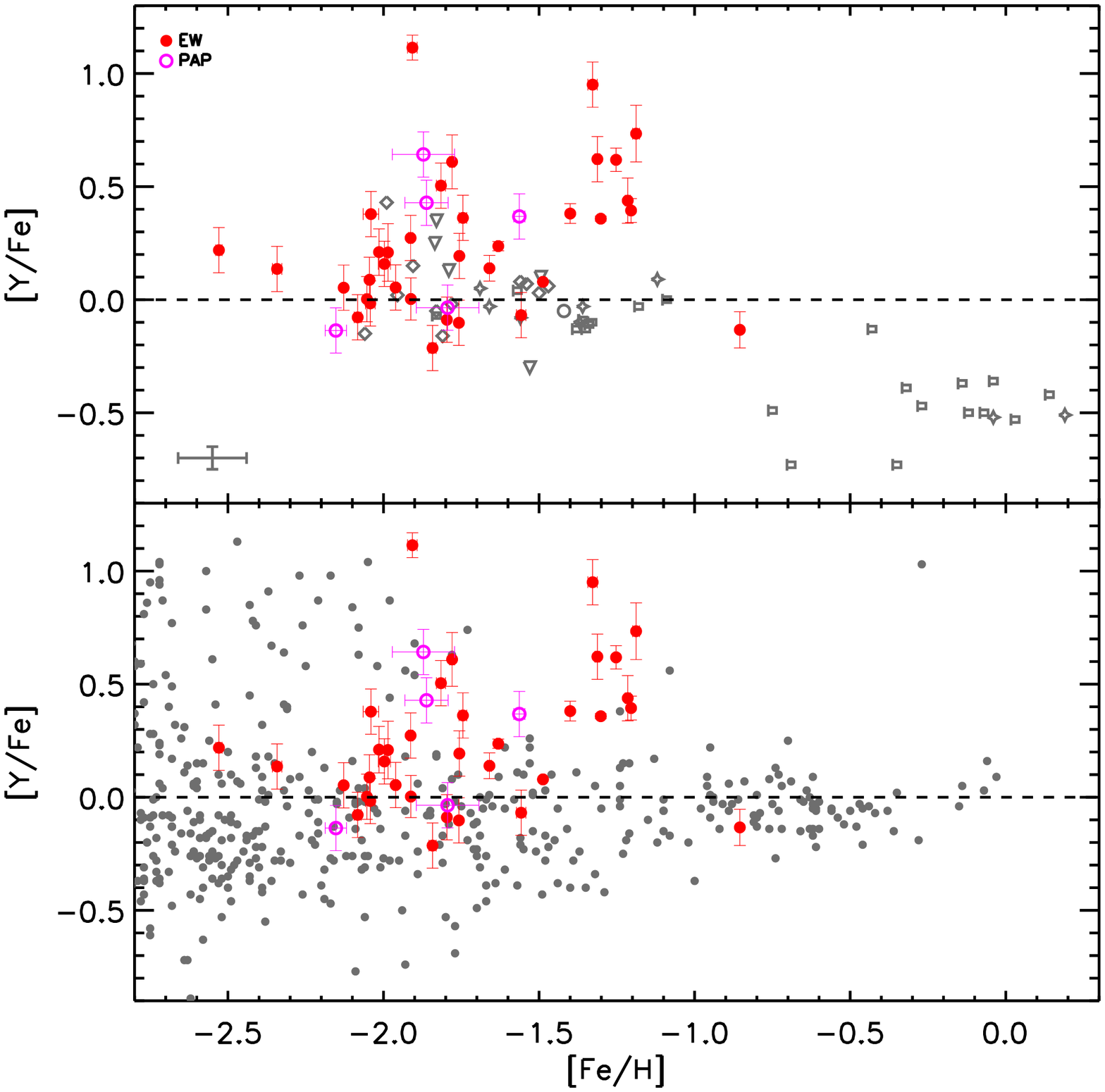}
\caption{Top panel: yttrium vs. iron abundances for the RRLs in \ocen. The stars analysed with the EW approach are marked with red filled circles, those analysed with the PAP approach are marked with magenta open circles. The black symbols mark the field halo RRLs collected with high-resolution spectroscopy from the literature. The black error bar on bottom left corner shows the mean individual errors for the literature sample. Bottom panel: as on top, but compared with field halo RGs \cite[black dots,][]{frebel10c}  \label{fig:ittrio}}
\end{figure}

\begin{figure}
\plotone{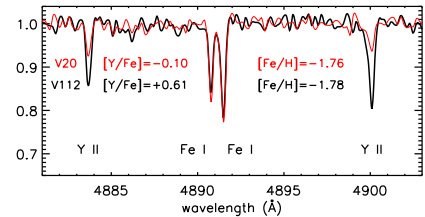}
\caption{Comparison of two spectra for similar RRLs (same pulsation mode, almost same phase and iron abundance) but with different Y enhancement.  \label{fig:ittrioenhanced}}
\end{figure}

Among the neutron-capture elements (Z~$>$~30), in the M2FS 
spectral range, only Y~{\sc ii} lines are easily observable 
in RRLs (Table~\ref{tab:elementi}).
A couple of La~{\sc ii} transitions are also present, but they are too 
weak to produce reliably detectable lines.
Indeed, synthetic spectrum tests show that these lines 
are not observable for [La/H]~$\lesssim$~$-$0.8, even with very high S/N. 
Y~{\sc ii} lines were measured in the entire
metallicity range covered by \ocen, with an average abundance for the entire sample
$\langle$[Y/Fe]$\rangle$~=~0.25~$\pm$~0.05 and a dispersion $\sigma$~=~0.31.
Figure~\ref{fig:ittrio} shows the comparison between the 
Y abundances of RRLs
in \ocen\ and the field halo RRLs (top panel, see Appendix in Paper~I for details) 
and RGs (bottom panel, \citealt{frebel10b}). 
Two groups of stars in \ocen\ show different levels of Y-enhancement:
about half of the sample is  in very good agreement with the field stars,
having about solar Y abundances;
the other half of the
RRLs shows a clear over-enhancement of Y, with [Y/Fe]~$\gtrsim$~0.4.
Figure~\ref{fig:ittrioenhanced} shows the comparison of two spectra: 
a Y-enhanced RRL (V112, [Y/Fe]~=~$+$0.61, black line) 
and an almost solar one (V20, [Y/Fe]~=~$-$0.10, red line). 
The two stars are both RRab, observed
at similar phase ($\phi$~=~0.28 and 0.36, respectively), 
and with similar iron abundance ([Fe/H]~=~$-$1.78 and $-$1.76, respectively).
The two shown iron lines are, indeed, almost identical, whereas the yttrium lines
are largely different one from each other.

The Y-enhanced group of RRLs
appears to be mostly the metal-rich one in 
\ocen\ ([Fe/H]~$\gtrsim$~$-$1.5), suggesting differential enrichments for 
two groups of RRLs.
A few other objects with similar strong Y-enhancement 
are also observed at lower metallicity ([Fe/H]~$\sim$~$-$1.8). However, they 
are a minor fraction of the RRLs more metal poor than [Fe/H]~$\lesssim$~$-$1.5.
The abrupt increase in the s-process element abundances with increasing [Fe/H]
was first observed by \cite{lloydevans83} and later confirmed by \cite{francois88},
\cite{paltoglou89}, and \cite{vanture94},  not only for Y, but also for La, Zr, Ba, and Nd.
\cite{johnson10} estimated the La abundance for $\sim$800 RGB stars, finding
a clear separation between the most metal-poor stars, with almost zero enhancement ([Fe/H]~$\lesssim$~$-$1.6, [La/Fe]~$\simeq$~0.0), and the most metal-rich, La
enhanced ([Fe/H]~$\gtrsim$~$-$1.6, [La/Fe]~$\simeq$~0.4).
The hypothesis by \cite{smith00} and \cite{cunha02} is that two different populations
coexist in \ocen, whose enrichment
history was strictly related to their capability to retain the products of the low velocity 
ejecta of asymptotic giant branch (AGB) stars wind
(rich in s-process elements), allowing a heavy self-enrichment of the 
second, metal-rich, stellar generation, over time scales of the order of 1~Gyr.
We note that the two groups of RRLs, the solar-enhanced and the 
over-enhanced, show similar radial distributions from the cluster center and similar kinematic 
properties (radial velocity, radial velocity dispersion). However, more 
statistics is required before we can reach a firm conclusion.

A different scenario was advanced by \cite{romano07}, 
who suggested that the self-enrichment scenario is not able to
reproduce the metallicity distribution of \ocen, and that the best 
hypothesis is that of \ocen\ as
the remnant of a dwarf spheroidal galaxy, evolved in isolation and then accreted by the Milky Way.
In favour of the working hypothesis suggested by 
\citeauthor{romano07}, let us mention that the 
Y-enhanced RRLs appear to be an isolated group 
in terms of Fe and Y abundances, i.e. the current data do not suggest a 
steady increase in Y when moving from metal-poor to metal-rich RRLs.  
It is also worth mentioning that \cite{norris96} suggested, on the basis of a large 
sample of Ca abundances of \ocen\ red giants,
that \ocen\ might be the merging of two different globulars.
This kind of enrichment in s-process elements has never 
been observed in other Galactic RRLs, and indeed, 
field RRLs do not show similar Y over-abundances.

Investigations of field stars and globular clusters in dwarf galaxies 
suggest that only a very limited
difference with respect to the Milky Way exists for Y and other s-process element abundances,
showing almost zero enhancement with respect to the Sun \citep[][and references therein]{tolstoy09}. 
Ba abundance in the Fornax dwarf galaxy represents a remarkable exception.
\cite{letarte10} found [Ba/Fe]~$\simeq$~0.7 for the investigated stars, 
more metal-rich than [Fe/H]~$\simeq$~$-$1.0. 
However, no similar enhancement was found for Y, and the problem
is still open.
Even if we can not easily distinguish two separate populations from the [Fe/H]
data, as was for S06 and B19, our [Y/Fe] abundances confirm that two distinct populations,
one more metal-poor and the other more metal-rich than [Fe/H]~$\simeq$~$-$1.5,
coexist in \ocen.
However, over-enhanced RRLs are also observed at [Fe/H]~$\sim$~$-$1.8,
and the most metal-rich RRL in our sample shows almost solar Y abundance, so that the distinction 
between the two metallicity groups is not strict.

\section{Conclusion and final remarks} \label{sec:fine}

We performed a large investigation of RR Lyrae stars in the globular
cluster \ocen, using high-resolution, high S/N spectroscopy.
We almost doubled the current sample of optical high-resolution 
spectroscopic abundances of RRLs, adding 109 cluster stars,
observed with M2FS at the {\it Magellan}/Clay Telescope,
to the $\sim$140 field halo stars available in the literature.

\ocen\ was confirmed as a complex cluster, with a broad metallicity range
and multiple populations.
Indeed, the samples of proprietary M2FS data and archive GIRAFFE data
allowed us to estimate $\langle$[Fe/H]$\rangle$~= $-$1.80~$\pm$~0.03,
with a high dispersion $\sigma$~=~0.33.
However, the average cluster metallicity alone is not sufficient to describe its complex nature.
In agreement with previous investigations of various
\ocen\ samples, we found a non-symmetric distribution of Fe, with a peak 
at [Fe/H]~$\simeq$~$-$1.85 and extended tails both in the metal-poor 
and especially in the metal-rich regime.
The peak of the distribution is $\sim$0.2~dex more metal-poor than previous estimates
for the cluster, with the exception of the work by \cite{bono19} who found
an even more metal-poor distribution.
The  $\alpha$- (Mg, Ca, and Ti) and iron-peak (Sc, Cr, Ni, and Zn) 
elements investigated show similar chemical enrichments to other 
known globular clusters and field stars of similar 
metallicity.
In particular, the agreement was found not only with RRL stars, 
as the ones in our sample, but in general with variable and non-variable 
field halo stars (RHB, BHB, and RGB stars), 
thus suggesting similar enrichment histories for all the analysed
old halo components.
The $\alpha$-elements are slightly enhanced, as expected for old stars,
with [$\alpha$/Fe]~=~0.41~$\pm$~0.02.
The iron-peak elements show almost solar abundances, with the exception of Zn 
that appears slightly enhanced.
On the contrary, the s-process element Y abundance shows peculiar characteristics, 
suggesting that two distinct populations coexist in the cluster, 
with the more metal-rich tail ([Fe/H]~$\gtrsim$~$-$1.5) dominated by stars with a strong
enhancement of s-process elements, well represented by the average abundance of
[Y/Fe]~$\gtrsim$~0.4, and the more metal-poor stars with almost solar abundance.
This over-enhancement of the metal-rich population has no comparison
in the field halo RRLs, appearing to be a peculiar characteristic of \ocen.

The cluster radial velocity was estimated with the help of multi-epoch
observations and template velocity curves
to remove the phase-to-phase variability due to pulsation for the individual observations.
We finally estimated the average velocity of \ocen\ as 
231.8~$\pm$~0.5~$\pm$~13.9~\kmsec,
in perfect agreement with literature results \citep{reijns06,an17}.

\acknowledgments

C.S. was partially supported by NSF grant AST-1616040, and 
by the Rex G. Baker, Jr. Endowment at the University of Texas.

C.S. also thanks the Dipartimento di Fisica -- Universit{\`a} di Roma Tor Vergata 
for a Visiting Scholar grant, and INAF -- Osservatorio Astronomico di Roma for his support during his stay.

We thank the reviewer for the precise notes and the useful suggestions to improve our work.

This research has made use of the services of the ESO Science Archive Facility.

This work has made use of the VALD database (\url{http://vald.astro.univie.ac.at/~vald3/php/vald.php}), 
operated at Uppsala University, the Institute of Astronomy RAS in Moscow, and the University of Vienna.

\appendix

\section{The non-RRL Stars In \ocen} \label{sec:app2}

For ten stars in our M2FS sample, either the spectra are significantly different from those 
expected for a RRL or the EW analysis produced equilibrium 
atmospheric parameters (\teff, \logg, \vmicro) that are not typical of RRLs.
However, their radial velocities confirm that these stars are actual members of \ocen.
The hypothesis is that the wrong stars were observed at the telescope due to
the crowding of the \ocen\ central region. Since we are not able to uniquely identify
these stars within the cluster, we name them as UNK (unknown), followed by a
sequential number corresponding to the RRL that was supposed to be observed 
(e.g. UNK15 was supposed to be the RRL V15 in \ocen).
We report in Table~\ref{tab:appendix1} a brief summary of their essential
atmospheric parameters and abundances.
As the nature for these objects is uncertain,
we report our results on them only for completeness, 
but we would recommend further investigations/observations 
before using them for scientific purposes. 

\begin{splitdeluxetable}{lccccBcccccccc}
\tablecaption{Parameters and abundances for the unknown, non-RRL stars. \label{tab:appendix1}}
\tablewidth{0pt}
\tablehead{
\colhead{ID} &
\colhead{\teff} &
\colhead{\logg} &
\colhead{\vmicro} &
\colhead{[Fe/H]} &
\colhead{[Mg/Fe]} &
\colhead{[Ca/Fe]} &
\colhead{[Ti/Fe]} &
\colhead{[Sc/Fe]} &
\colhead{[Cr/Fe]} &
\colhead{[Ni/Fe]} &
\colhead{[Zn/Fe]} &
\colhead{[Y/Fe]} 
}
\startdata
UNK15  & 8200 &  4.30 & 4.50  &$-$1.21  $\pm$    0.01 &      0.19                &   \ldots                &    0.61  $\pm$    0.04 &   \ldots                &   \ldots                &  \ldots                &   \ldots                 & \ldots                 \\
UNK19  & 7100 &  4.00 & 5.10  &$-$1.49  $\pm$    0.02 &    \ldots                &   \ldots                &    1.09  $\pm$    0.14 &   \ldots                &   \ldots                &  \ldots                &   \ldots                 &   1.12                 \\
UNK90  & 5700 &  3.60 & 2.00  &$-$1.15  $\pm$    0.04 &   $-$0.41                &   \ldots                &    0.66  $\pm$    0.08 &     0.44  $\pm$    0.23 &     0.21  $\pm$    0.10 &    0.03  $\pm$    0.07 &     0.22                 &   0.61  $\pm$    0.10  \\
UNK109 & 7500 &  3.90 & 2.50  &$-$0.92  $\pm$    0.04 &      0.16                &   \ldots                &  \ldots                &     0.38                &   \ldots                &  \ldots                &   \ldots                 & \ldots                 \\
UNK114 & 5600 &  4.10 & 2.00  &$-$2.12  $\pm$    0.07 &    \ldots                &     0.64                &    0.90  $\pm$    0.17 &     0.74                &     0.45  $\pm$    0.50 &    0.23  $\pm$    0.26 &     0.73                 &   0.78  $\pm$    0.21  \\
UNK118 & 6900 &  4.90 & 0.60  &$-$0.59  $\pm$    0.03 &   $-$0.71                &   \ldots                &    0.72  $\pm$    0.28 &     0.30                &     0.09  $\pm$    0.08 &  \ldots                &   \ldots                 & \ldots                 \\
UNK143 & 5800 &  3.00 & 3.00  &$-$2.08  $\pm$    0.04 &      0.21                &   \ldots                &    0.89  $\pm$    0.08 &     0.39                &     0.58                &  \ldots                &   \ldots                 &   0.48                 \\
UNK146 & 5300 &  0.20 & 2.25  &$-$2.21  $\pm$    0.01 &      1.72                &   \ldots                &    0.22  $\pm$    0.18 &   \ldots                &  $-$0.20                &  \ldots                &   \ldots                 &$-$0.07  $\pm$    0.10  \\
UNK267 & 4900 &  0.90 & 2.60  &$-$3.01  $\pm$    0.05 &    \ldots                &     1.48                &    0.59  $\pm$    0.02 &   \ldots                &     0.71  $\pm$    0.04 &  \ldots                &   \ldots                 &$-$0.12                 \\
UNK277 & 5800 &  3.00 & 1.80  &$-$1.40  $\pm$    0.05 &    \ldots                &   \ldots                &    0.38  $\pm$    0.12 &   \ldots                &     0.19  $\pm$    0.08 &    0.09  $\pm$    0.16 &  $-$0.09                 &   0.23                 \\
\enddata
\end{splitdeluxetable}

\bibliographystyle{aasjournal}
\bibliography{ms}

\end{document}